\documentclass[letter,useAMS,usenatbib]{mn2e}
\usepackage[pdftex]{graphicx}

\def\msun{M$_\odot$}
\def\mstar{$M_\star$}

\def\mic{$\mu$m}
\def\arcsec{$^{\prime\prime}$}

\def\ha{H$\alpha$}
\def\hb{H$\beta$}
\def\oii{[OII]$\lambda$3727}
\def\nii{[NII]$\lambda$6584}

\def\oiiib{[OIII]$\lambda$5007}
\def\oiii{[OIII]$\lambda$4958,5007}

\title[The Fundamental Metallicity Relation]
{A fundamental relation between mass, SFR and metallicity in local 
and high redshift galaxies}

\author[F. Mannucci et al.]{
F. Mannucci$^1$\thanks{E-mail:filippo@arcetri.astro.it},
G. Cresci$^{1,2}$,
R. Maiolino$^{3}$,
A. Marconi$^{4}$,
A. Gnerucci$^{4}$\\
$^1$INAF - Osservatorio Astrofisico di Arcetri, 
   Largo E. Fermi 5, I-50125, Firenze, Italy\\
$^2$Max-Planck-Institut f\"ur extraterrestrische Physik (MPE), 
   Giessenbachstr.1, D-85748 Garching, Germany\\
$^3$INAF - Osservatorio Astronomico di Roma, via di Frascati 33, 
   I-00040 Monte Porzio Catone, Italy\\
$^4$Dip. di Fisica e Astronomia, Universit\`a di Firenze, 
   Largo E. Fermi 2, I-50125, Firenze, Italy\\
}

\begin{document}

\date{Submitted 2010 March}

\pagerange{\pageref{firstpage}--\pageref{lastpage}} \pubyear{2010}

\maketitle

%-------------------------------------------------------------------------
\begin{abstract}
We show that the mass-metallicity relation observed in the local universe 
is due to a more general relation between stellar mass \mstar, 
gas-phase metallicity and star formation rate (SFR). 
Local galaxies define a tight surface in this 3D space, the 
Fundamental Metallicity Relation (FMR), with a 
small residual dispersion of $\sim$0.05~dex in metallicity, i.e, $\sim$12\%.
At low stellar mass, metallicity decreases sharply with increasing SFR,
while at high stellar mass, metallicity does not depend on SFR.

High redshift galaxies, up to z$\sim$2.5 are found to follow the 
same FMR defined by local SDSS galaxies,
with no indication of evolution. In this respect, 
the FMR defines the properties of metal enrichment of galaxies 
in the last 80\% of cosmic time.
The evolution of the mass-metallicity relation observed up to z=2.5
is due to the fact that galaxies with 
progressively higher SFRs, and therefore lower metallicities, 
are selected at increasing redshifts,
sampling different parts of the same FMR. 

By introducing the new quantity $\mu_\alpha=\rm{log}(M_\star)-\alpha\,\rm{log}(SFR)$,
with $\alpha$=0.32,
we define a projection of the FMR that minimizes the metallicity scatter
of local galaxies.  The same quantity also cancels out any redshift 
evolution up to z$\sim$2.5, i.e, all galaxies follow the same relation 
between $\mu_{0.32}$ and metallicity and have the same range of values of $\mu_{0.32}$. 
At z$>$2.5, evolution of about 0.6~dex off the FMR is observed,
with high-redshift galaxies showing lower metallicities.

The existence of the FMR can be explained by the interplay of infall of pristine gas
and outflow of enriched material. The former effect is responsible for the dependence of 
metallicity with SFR and is the dominant effect at high-redshift, 
while the latter introduces 
the dependence on stellar mass and dominates at low redshift. The combination
of these two effects, together with the Schmidt-Kennicutt law, 
explains the shape of the FMR and the role of $\mu_{0.32}$. 
The small metallicity scatter around
the FMR supports the smooth infall scenario of gas accretion in the local universe.

\end{abstract}

\begin{keywords}
Galaxies: abundances; Galaxies: formation; 
Galaxies: high-redshift;Galaxies: starburst
\end{keywords}
%
%=========================================================================
\section{Introduction}
\label{sec:intro}

Gas metallicity is regulated by a complex interplay between 
star formation, infall of metal-poor gas and outflow of enriched material. 
A relation between magnitude and metallicity was discovered 
in the '70 \citep{McClure68,Lequeux79}, in which more luminous galaxies also have 
higher metallicities. 
Later on, it was understood that this luminosity-metallicity relation
is a manifestation of a more fundamental stellar mass-metallicity relation
where galaxies with larger stellar mass \mstar\  
have higher metallicities
\citep{Garnett02,Perez-Gonzalez03a,Pilyugin04,Tremonti04,Savaglio05,Lee06,
Cowie08,Panter08,Kewley08,Hayashi09a,Michel-Dansac08,Liu08,Rodrigues08,
Perez-Montero09}.

The origin of this relation is debated, and many different explanations
have been proposed, including ejection of metal-enriched gas
(e.g., \citealt{Edmunds90,Lehnert96a,Garnett02,Tremonti04,Kobayashi07,Scannapieco08,Spitoni10}),
``downsizing'', i.e., a systematic dependence of the efficiency of star 
formation with galaxy mass  
(e.g., \citealt{Brooks07,Mouchine08,Calura09a}), variation of the IMF
with galaxy mass \citep{Koppen07}, and infall of metal-poor gas
\citep{Finlator08,Dave10}.

Recently, evidence has been found that at high redshift
the infall of pristine gas can have a dominant role
\citep{Bournaud09,Dekel09,Brooks09,Agertz09b}. 
The mass-metallicity relation has been studied
by \cite{Erb06a}  at z$\sim$2.2 and by
\cite{Maiolino08} and \cite{Mannucci09b} at z=3--4, finding 
a strong and monotonic evolution, with metallicity decreasing with redshift
at a given mass.
The same authors \citep{Erb06a,Erb08,Mannucci09b} 
have also studied the relation between metallicity and gas fraction, i.e., 
the effective yields, obtaining clear 
evidence of the presence of infall in high redshift galaxies. 
% This seems to be in contrast with the local universe, 
% where the mass-metallicity relation
% seems to be driven by outflows \citep{Tremonti04,Spitoni10}.

Even more recently, \cite{Cresci10} have
obtained the evidence of ``positive'' metallicity gradients in a small 
sample of three disk galaxies at z=3.5, with lower metallicities in more active 
regions of star formation. The presence of these gradients in galaxies with 
very regular dynamics
can be understood as a consequence of 
accretion of metal-poor gas, producing a new episode of star- formation
in older, more metal-rich galaxies.

If infall is at the origin of the star formation activity,
and outflows are produced by exploding supernovae (SNe),
a relation between metallicity and SFR
is likely to exist. In other words, SFR is a parameter
that should be considered in the scaling relations that include metallicity.
The role of specific SFR (SSFR) in this content was already studied by 
\cite{Ellison08a},
who presented a mild ($\le$0.1~dex) dependence of metallicity on SSFR
when binning galaxies according to their \mstar. 
Recently, also 
\cite{Lopez-Sanchez10c} presented evidence for a link between SFR and metallicity,
while \cite{Peeples09} reported high values of SFR in a sample of
outliers, toward low metallicities, of the mass-metallicity relation.

To test the hypothesis of a correlation between SFR and metallicity
in the present universe and at high redshift,
we have studied 
several samples of galaxies at different redshifts whose 
metallicity, \mstar, and SFR have been measured.
In the next section we present the data samples we are using for this study. 
In sec.~\ref{sec:massmet} we study the mass-metallicity 
relation as a function of SFR, and in the following section
we introduce the Fundamental Metallicity Relation.
In sec.~\ref{sec:discussion} we discuss the physical origin of this relation.
We adopt a $\Lambda$CDM cosmology with 
$H_0$=70 km/sec, $\Omega_m$=0.3 and $\Omega_\Lambda$=0.7. Stellar mass \mstar\ and SFR
are expressed in \msun\ and in \msun/yr, respectively.\\

%---------------------------------------------------------------
\begin{figure*}
\centerline{
   \includegraphics[width=0.48\textwidth]{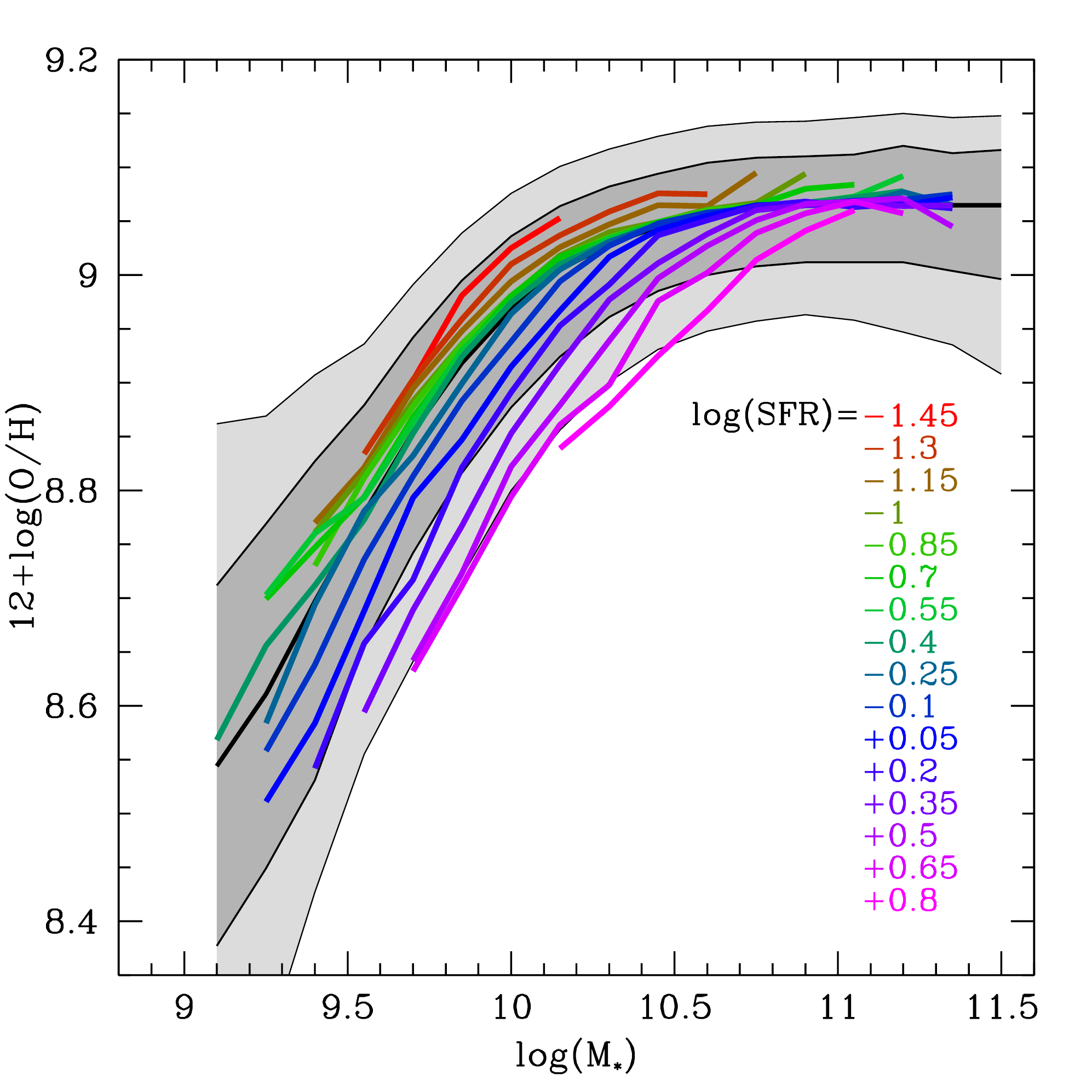} 
   \includegraphics[width=0.48\textwidth]{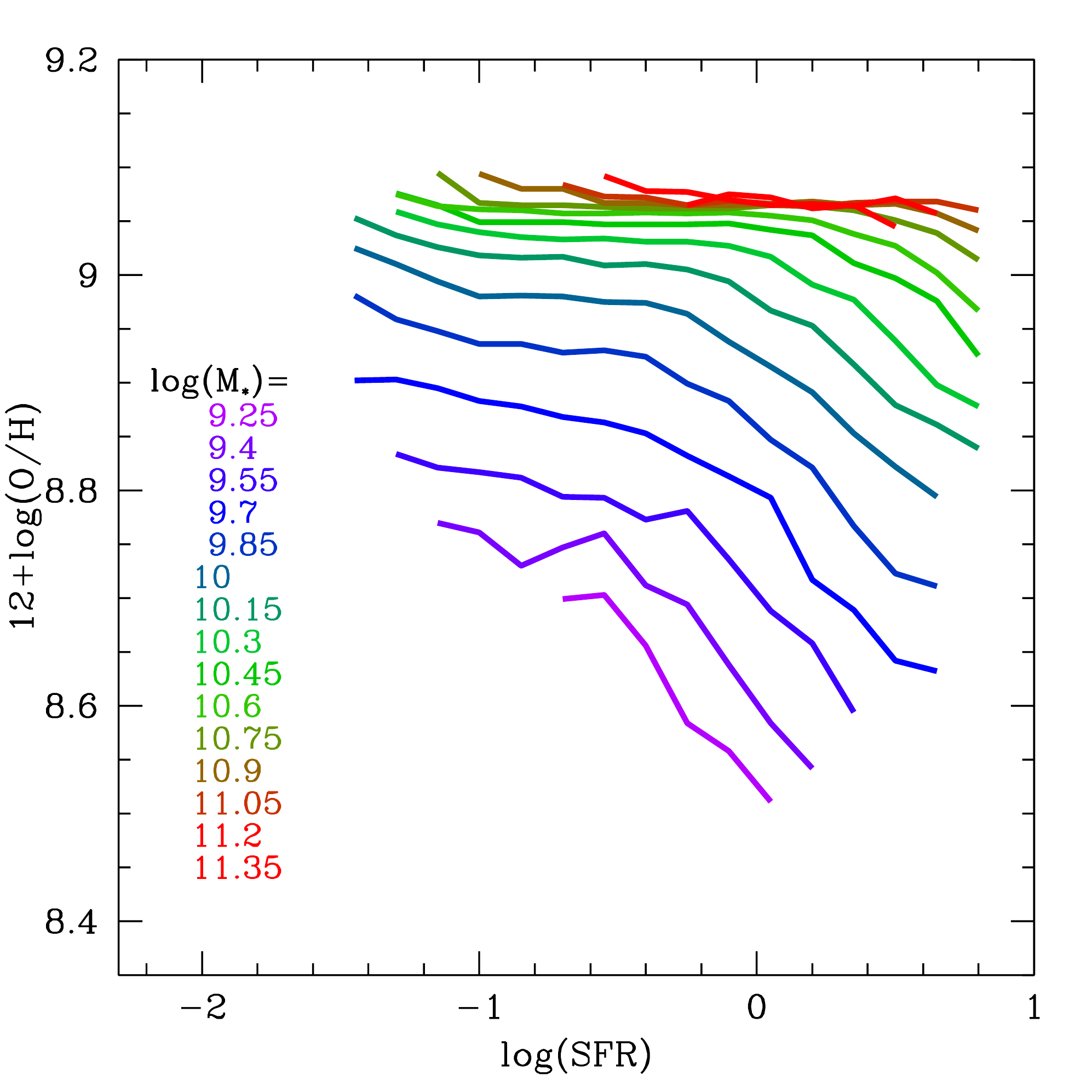} 
}
\caption{
{\em Left panel:} The mass-metallicity relation of local SDSS galaxies. 
The grey-shaded areas contain 64\% and 90\% of all SDSS galaxies, with the 
thick central line showing the median relation. The colored lines show the 
median metallicities, as a function of \mstar, 
of SDSS galaxies with different values of SFR. 
% The yellow line shows the 
% polynomial best fit to the median metallicity.
{\em Right panel:} median metallicity as a function of SFR for galaxies of 
different \mstar.
At all \mstar\ with log(\mstar)$<$10.7,
metallicity decreases with increasing SFR 
at constant mass .
}
\label{fig:massmet}
\end{figure*}

%==============================================================================
\section{The galaxy samples}

\subsection{z=0: SDSS}
\label{sec:sdss}

Local galaxies are well measured by the SDSS project
\citep{Abazajian09}. We used the MPA/JHU
catalog of emission line fluxes and stellar masses from SDSS-DR7 available at
http://www.mpa-garching.mpg.de/SDSS
and described in \cite{Kauffmann03a}, \cite{Brinchmann04} and \cite{Salim07}.
The catalog includes 927552  galaxies whose spectroscopic properties,
such as emission line fluxes and spectroscopic indexes,
have been measured.

We selected emission-line galaxies with redshift between 0.07 and 0.30 
(47\% of the total sample).
The minimum redshift is set in order to
ensure that \oii\ is well within the useful spectral range, and that
the 3\arcsec\ aperture of the spectroscopic fiber samples a significant 
fraction of the galaxies (3\arcsec\ correspond to $\sim$4 kpc at z=0.07).

A threshold to the signal-to-noise ratio (SNR) of \ha\ 
of SNR$>$25 was used to have reliable values of metallicity,
while no SNR threshold is used for the other lines.
Such a high SNR on \ha\ is needed to ensure that all the main optical lines 
are generally detected with enough SNR without introducing metallicity biases.
For example, the \nii\ flux is about 1/2 of the \ha\  flux at high metallicities, 
and about 1/10 at the lowest metallicities sampled by SDSS. 
Noise and intrinsic dispersion of the \nii/\ha\ ratio could produce
very low fluxes of the \nii\ in low metallicity galaxies. 
If a threshold in the SNR of \nii\ is used, some low-metallicity galaxies
would be removed from the sample, while this effect would be negligible
among high-metallicity galaxies. As a result
a metallicity bias would be introduced, and such a bias can be important when 
high SNR thresholds are used. Also,
this selection bias would be correlated
with SFR as more active galaxies have brighter lines.
The use of a SNR threshold of 25 on \ha\ means that, on average, 
the faintest \nii\ lines are detected with SNR$>$2.5, but even
galaxies with lower SNR for this line are included in the sample. 
This limit on SNR selects 43\% of the remaining sample. 

We limited dust extinction
to $A_V<2.5$, in order not to deal with very large 
extinction corrections, and galaxies with Balmer decrements 
below 2.5 were removed. These selections remove 0.2\% of the galaxies. 
Finally, 
AGN-like galaxies (22\% of the sample) were excluded by using the BPT classification
by \cite{Kauffmann03c}.

Total stellar masses \mstar\  from \cite{Kauffmann03a} were used, as listed 
in the same MPA/JHU catalog, with a correction factor of 1.06 
to scale the masses down from a \cite{Kroupa01} to a \cite{Chabrier03} 
initial mass function (IMF).

SFRs inside the spectroscopic aperture were measured 
from the \ha\ emission line flux corrected for dust extinction 
as estimated from the Balmer decrement.
The conversion factor between \ha\ luminosity and SFR 
in \cite{Kennicutt98} was used,
corrected to a \cite{Chabrier03} IMF.

Oxygen gas-phase abundances were measured from the emission line ratios
as described in \cite{Nagao06} and \cite{Maiolino08}. Two independent 
measurements of metallicity are available for these galaxies, 
based either on the \nii/\ha\ ratio or on the R23 quantity, defined as
R23=(\oii+\oiii)/\hb. 
When both quantities are inside the useful range for metallicity calibration,
i.e., log(\nii/\ha)$<$--0.35 and log(R23)$<$0.90, we selected only
galaxies where the two values of metallicity differ less that 0.25~dex 
(97\% of the sample),
and galaxy metallicity is then defined as the average of these two values.
The spread of the difference between these two estimates of metallicity 
is 0.09~dex ($\sim$23\%),
with a significant albeit small systematic difference of 0.05~dex ($\sim$12\%) 
with the value from R23 systematically higher than that derived from \nii/\ha.
This small difference is likely to be due to the different sample used here and 
in \cite{Maiolino08},
which use a SNR threshold of 10 on the flux of each line.
This may introduce a small bias in the calibrations relative to our sample.

The final galaxy sample contains 141825 galaxies.

%----------------------------------------------------------------------------
\subsection{z=0.5--2.5}
\label{sec:z2}

Many galaxies has been observed
at high redshift and these data can be used to study the evolution of 
metallicity with respect to the other properties of galaxies.
We extracted from the literature three samples of galaxies
at intermediate redshifts, for a total of 182 objects, 
having published values of emission line fluxes, \mstar, and
dust extinction:
 0.5$<$z$<$0.9 (\citealt{Savaglio05}, GDDS galaxies), 
 1.0$<$z$<$1.6 \citep{Shapley05a,Liu08,Wright09,Epinat09a}, and 
 2.0$<$z$<$2.5 \citep{Law09b,Lehnert09,Forster-Schreiber09}. 
The same procedure used for the SDSS galaxies was applied
to these galaxies. 
Metallicity is estimated either from R23 or from \nii/\ha, depending on which 
lines are available. AGN are removed using the BPT diagram \citep{Kauffmann03c}
or, when \oiiib\ and \hb\ are not available, by imposing 
log(\nii/\ha)$<$--0.3.
The \nii\ line, which is usually much fainter than \ha, is 
not detected in several galaxies, but removing these galaxies from the sample
would bias it towards high metallicities. 
For these objects we have assumed a value of the intrinsic 
\nii\ flux which is half of the upper limiting flux.
When necessary, the published \mstar\ have been converted to a 
\cite{Chabrier03} IMF.
For galaxies without observations of both \ha\ and \hb, dust extinction 
is estimated from SED fitting, and we assume that continuum and 
the emission lines suffer the same extinction. In local starburst 
lines often suffer of higher extinctions ($A_V$(lines)$\sim$2.3$A_V$(SED)
according to \citealt{Calzetti00}). We have checked that the inclusion of this effect 
would have little effect on the final relations and on the conclusions of this paper.

\cite{Erb06a} have observed a large sample of 91 galaxies
at z$\sim$2.2. Metallicities have been measured only on average spectra 
stacked according to \mstar, which has the results of mixing galaxies of 
different SFRs. Despite this problem, no systematic differences in metallicity 
are detected with respect to the other galaxies measured individually, 
and the \cite{Erb06a} galaxies are included in the high-redshift sample, 
although without binning them with the rest of the galaxies.

%---------------------------------------------------------------
\begin{figure*}
\centerline{
	\includegraphics[width=0.55\textwidth]{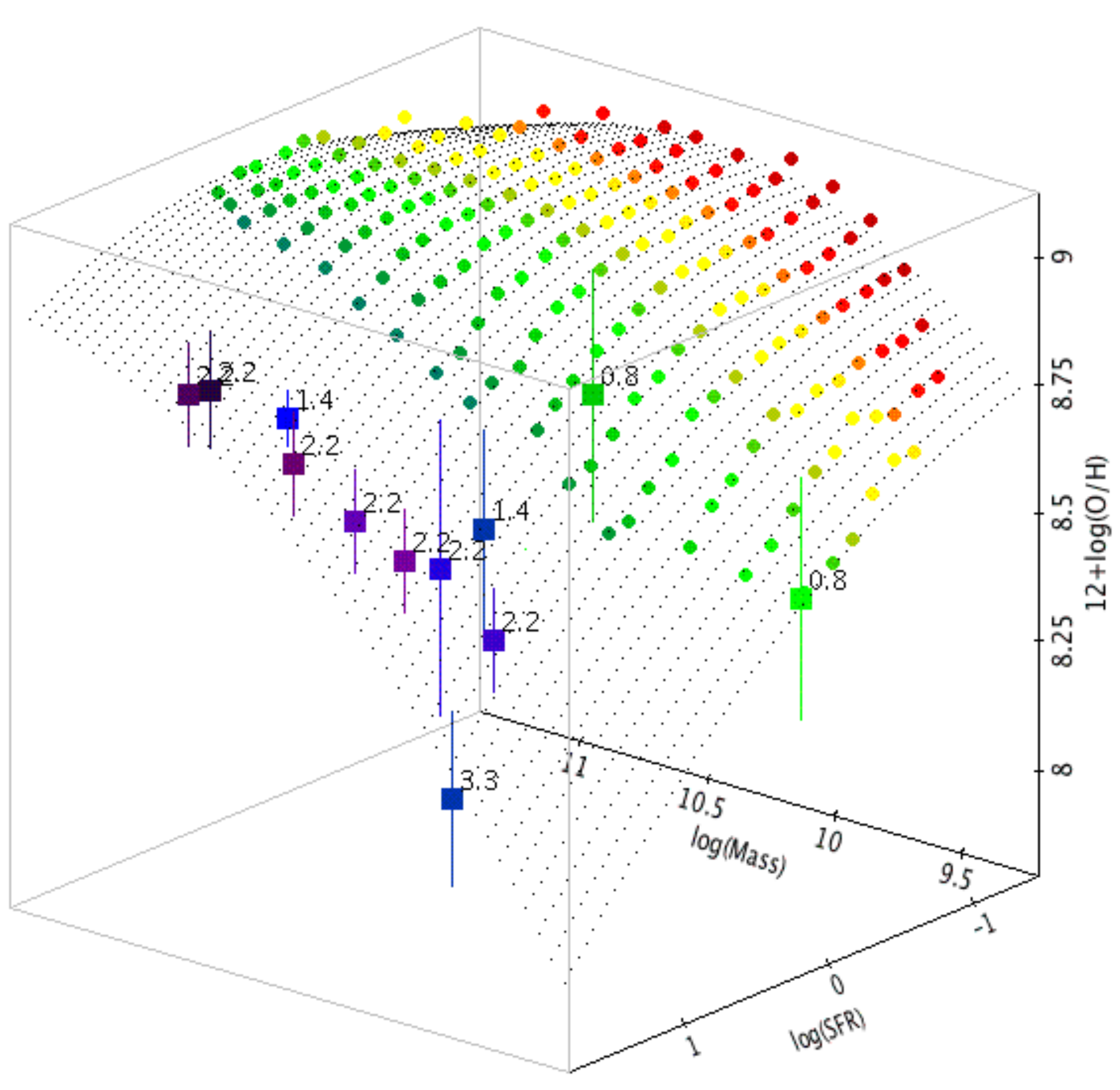} 
	\includegraphics[width=0.05\textwidth]{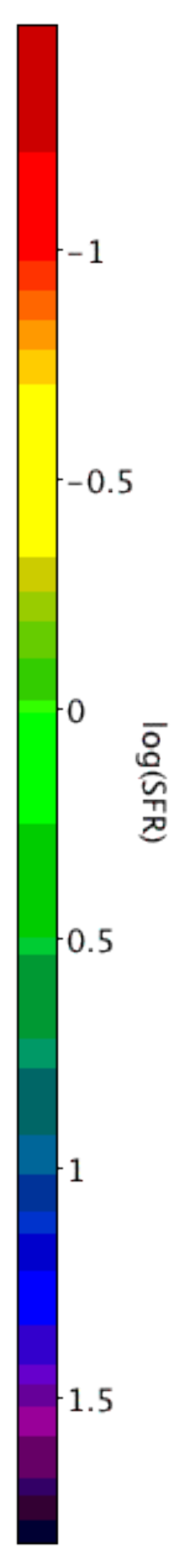} 
}
\centerline{
	\includegraphics[width=0.40\textwidth]{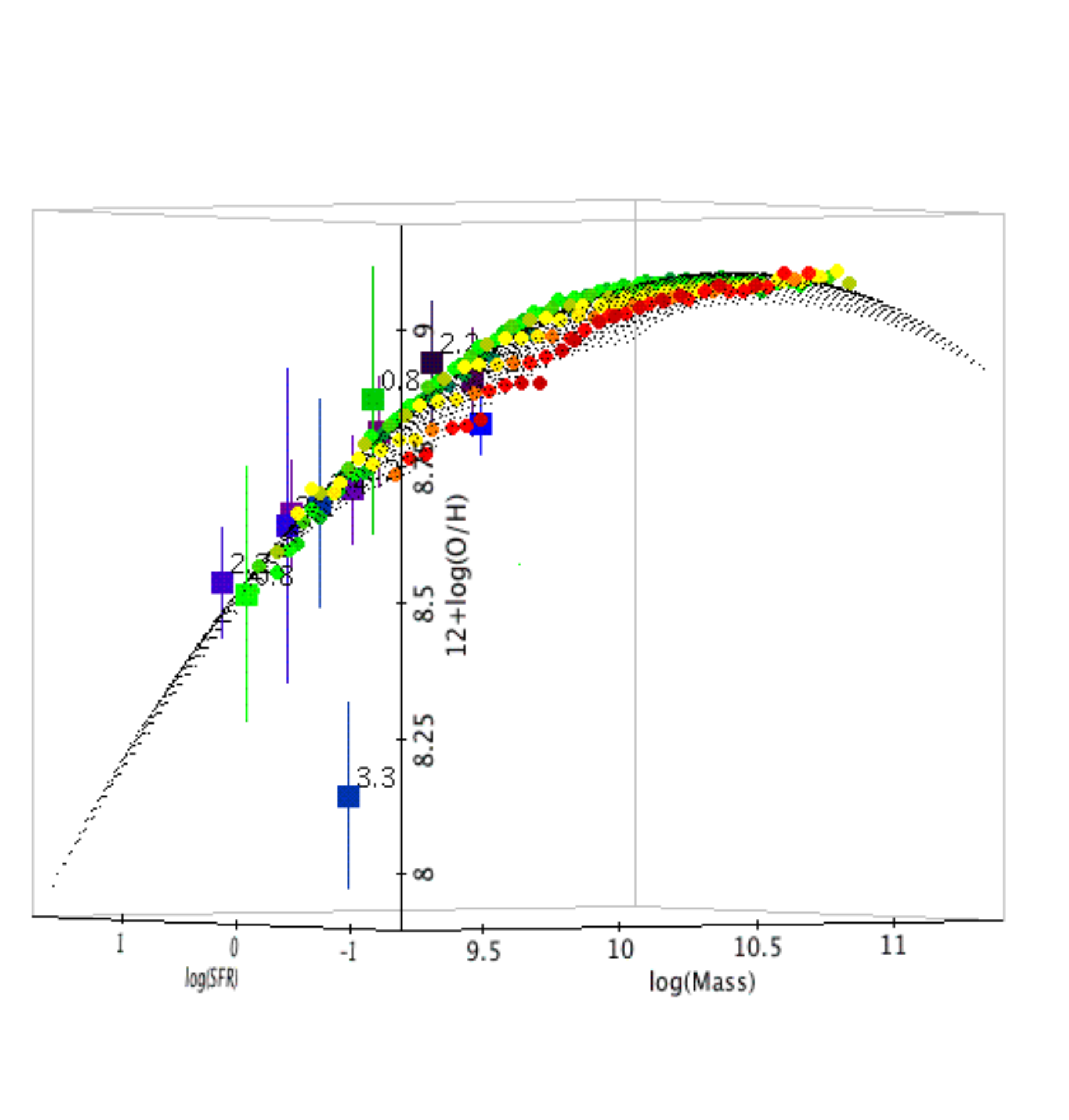} 
	\includegraphics[width=0.40\textwidth]{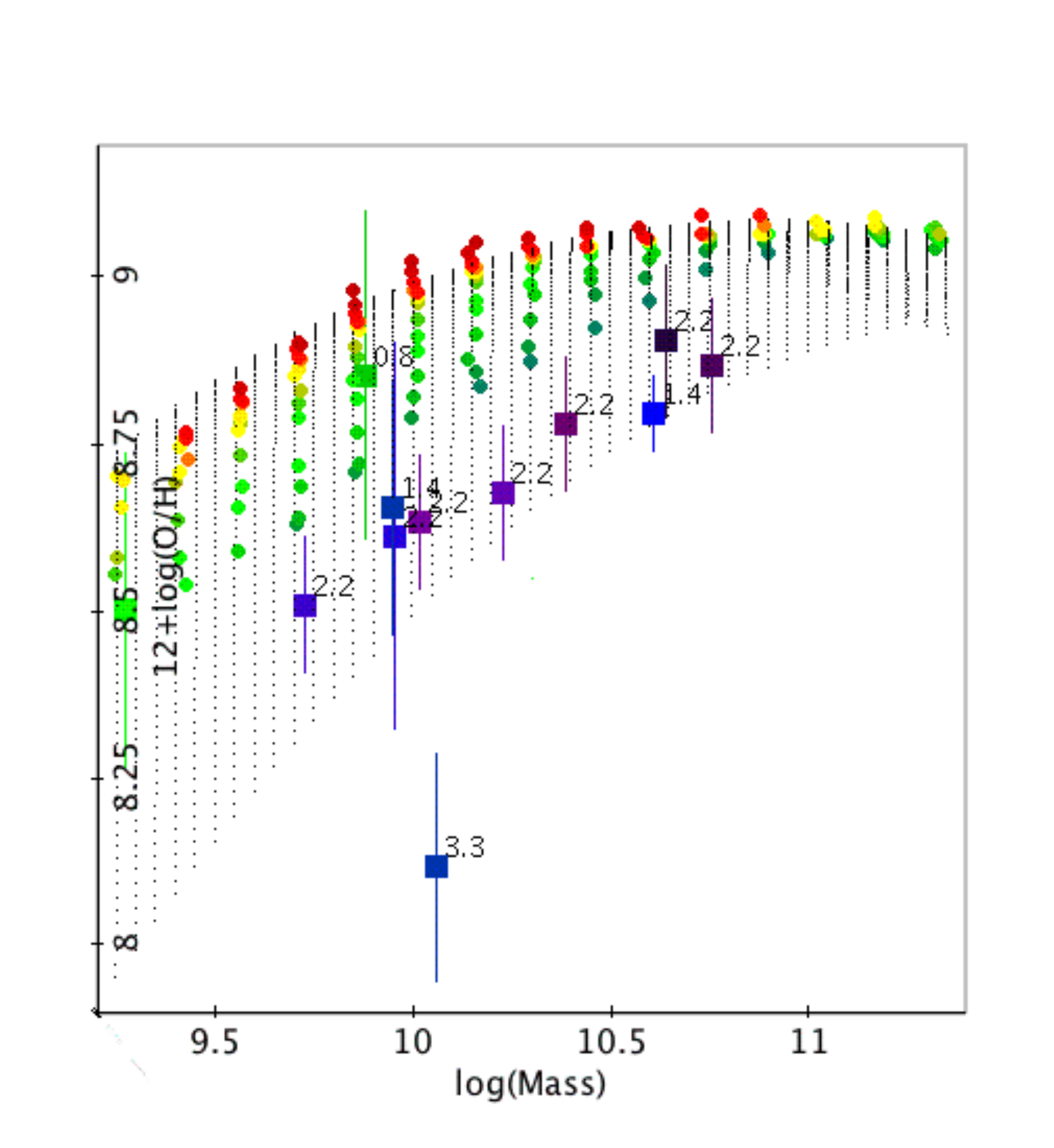} 
}
\caption{
Three projections of the Fundamental Metallicity Relation 
among \mstar, SFR and gas-phase metallicity. 
Circles without error bars are the median values of metallicity of local SDSS galaxies
in bin of \mstar\ and SFR, color-coded with SFR as shown in the colorbar on the right.
These galaxies define a tight surface in the 3D space, with
dispersion of single galaxies around this surface of $\sim$0.05~dex.
The black dots show a second-order fit to these SDSS data, 
extrapolated toward higher SFR.
Square dots with error bars are the median values of high redshift galaxies, 
as explained in the text.
Labels show the corresponding redshifts.
The projection in the lower-left panel emphasizes that most of the high-redshift data,
except the point at z=3.3, are found on the same surface defined by low-redshift data.
The projection in the lower-right panel corresponds to the mass-metallicity relation, as 
in Fig.~\ref{fig:massmet}, showing that the origin of the observed evolution in 
metallicity up to z=2.5 is due to the progressively increasing SFR.
}
\label{fig:cfr1}
\end{figure*}

%----------------------------------------------------------------------------
\subsection{z=3--4}
\label{sec:z3}

A significant sample of 16 galaxies at redshift between 3 and 4 was
observed by \cite{Maiolino08} and \cite{Mannucci09b} for the LSD 
and AMAZE projects.
Published values of stellar masses, line fluxes and metallicities
are available for these galaxies, which can be 
compared with lower redshift data.
The same procedure as at lower redshift was used, with the exception that
SFR is estimated from \hb\ after correction for dust extinction,
and metallicities are measured by a simultaneous fitting of the 
line ratios involving \oii, \hb\ and \oiii, as described in \cite{Maiolino08}.

%==============================================================================
\section{The mass-metallicity relation as a function of SFR}
\label{sec:massmet}

The grey-shaded area in the left panel of Fig.~\ref{fig:massmet} 
shows the mass-metallicity relation for our sample of SDSS galaxies. 
Despite the differences in the
selection of the sample and in the measure of metallicity, 
our results are very similar to what has been found by \cite{Tremonti04}.
The metallicity dispersion of our sample, $\sim$0.08~dex, is somewhat 
smaller to what
have been found by these authors, $\sim$0.10~dex, possibly due to different
sample selections and metallicity calibration.
The 4th-order polynomial fit to the median mass-metallicity relation is:
\begin{equation}
\label{eq:mmfit}
\begin{array}{rl}
12+log(O/H)=&8.96+0.31m-0.23m^2\\
            &-0.017m^3+0.046m^4\\
\end{array}
\end{equation}
where $m$=log(\mstar)--10 in solar units.

We have computed the median metallicity of SDSS galaxies for different 
values of SFR.  Median have been computed in bins of mass and SFR of 0.15~dex 
width in both quantities.
On average, each bin contains 760 galaxies, and only bins containing 
more than 50 galaxies are considered. 
The left panel of Fig.~\ref{fig:massmet} also shows these median metallicities
as a function of \mstar.
It is evident that a systematic segregation in SFR is present in the data.
While galaxies with high \mstar\ (log(\mstar)$>$10.9) show no correlation 
between metallicity and SFR, at low \mstar\ more active galaxies 
also show lower metallicity.
The same systematic dependence of metallicity on SFR can be seen in 
the right panel of Fig.~\ref{fig:massmet},
where metallicity is plotted as a function of SFR for different values of mass.
Galaxies with high SFRs show a sharp dependence of metallicity on SFR, while
less active galaxies show a less pronounced dependence.

A hint of this effect was already noted by \cite{Ellison08a}, but the 
different sample selection and the large bins in SFR reduced the observed 
dependence on
SFR to a small correction of $\sim$0.05~dex with respect to the value determined by
the mass-metallicity relation. Also \cite{Rupke08} presented evidences
for lower metallicities in local galaxies with high SFRs, although with 
a very large scatter.

%------------------------------------------------------------------------------
\section{The Fundamental Metallicity Relation}
\label{sec:cfr}

The dependence of metallicity on \mstar\ and SFR can be better visualized 
in a 3D space with these three coordinates, as shown in Figure~\ref{fig:cfr1}.
SDSS galaxies appear to 
define a tight surface in the space, the Fundamental Metallicity Relation,
with metallicity well defined by the values of \mstar\ and SFR.  
All the data on this FMR are shown in table~1.\\

\begin{figure}
\centerline{
	\includegraphics[width=0.9\columnwidth]{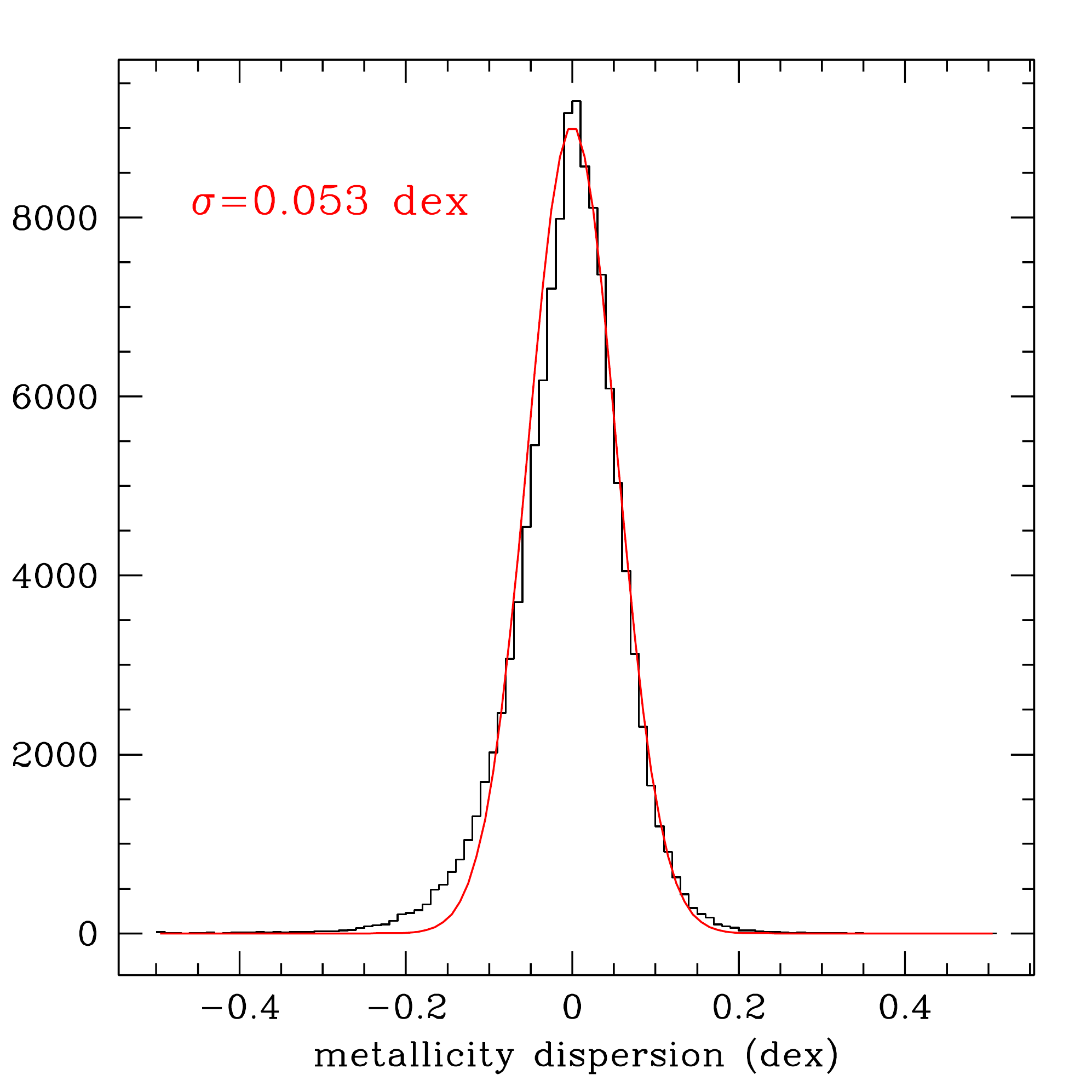} 
}
\caption{
Metallicity dispersion of single SDSS galaxies around the FMR. 
This histogram shows the 
differences from the median computed in bins of 0.05~dex in \mstar\ and SFR. 
The red line is a gaussian with $\sigma$=0.053~dex.
Positive differences mean metallicities of single galaxies larger 
than the median.
}
\label{fig:plotdif}
\end{figure}

The introduction of the FMR results in a significant reduction of residual 
metallicity scatter with respect to the simple mass-metallicity relation. 
The dispersion of individual SDSS galaxies around the FMR, 
shown in Fig.~\ref{fig:plotdif}, computed in bins 
of 0.05~dex in \mstar\ and SFR,
is $\sim$0.06~dex when computed across the full FMR and reduces to $\sim$0.05~dex
i.e, about 12\%, in the central part of the relation where most of the 
galaxies are found.
This means that about half 
of the total scatter of the mass-metallicity relation 
(0.08~dex) is due to the systematic effect with SFR,
while about half is due to intrinsic differences between galaxies
and/or uncertainties on the measurements. 
The reduction in scatter becomes even more significant when considering
that most of the galaxies in the sample cover 
a small range in SFR, with 64\% of the galaxies ($\pm$1$\sigma$) 
is contained inside 0.8~dex.
Galaxies with very low SFRs are not selected due to the high SNR threshold 
on \ha, while high SFR galaxies are rare in the local universe.
As a consequence
the scatter on the full sample is dominated by galaxies having a small
correction due to SFR.  In contrast, considering  only
galaxies with high SFRs, the scatter is reduced by a large factor. 
For example, for log(SFR)$>$0.5, 
the scatter around the average mass-metallicity relation is 
about 40\%, while it is a factor of $\sim$3 lower 
around the FMR.

%---------------------------------------------------------------
%---------------------------------------------------------------

\begin{table*}
\label{tab:fmr}
\caption{The Metallicity Fundamental Relation for SDSS galaxies
selected as in sec.~\ref{sec:sdss}.
For each value of \mstar\ (reported in the first line)
and SFR (first column)
we list the median value of metallicity, 
the 1$\sigma$ dispersion around this value, 
and the number of galaxies in each bin.}
\footnotesize
\renewcommand{\tabcolsep}{3pt}
\begin{tabular}{r|cccccccccccccccc}
\hline
\hline
log(SFR) & \multicolumn{16}{c}{log(\mstar)} \\
        &  9.10 &  9.25 &  9.40 &  9.55 &  9.70 &  9.85 & 10.00 & 10.15 & 10.30 & 10.45 & 10.60 & 10.75 & 10.90 & 11.05 & 11.20 & 11.35 \\ 
\hline
$-$1.45 &       &       &       &       &  8.90 &  8.98 &  9.02 &  9.05 &       &       &       &       &       &       &       &       \\ 
        &       &       &       &       &  0.08 &  0.07 &  0.06 &  0.07 &       &       &       &       &       &       &       &       \\ 
        &       &       &       &       &    88 &   116 &    94 &    68 &       &       &       &       &       &       &       &       \vspace{2pt}\\ 
$-$1.30 &       &       &       &  8.83 &  8.90 &  8.96 &  9.01 &  9.04 &  9.06 &  9.08 &  9.09 &       &       &       &       &       \\ 
        &       &       &       &  0.07 &  0.07 &  0.07 &  0.06 &  0.06 &  0.06 &  0.07 &  0.08 &       &       &       &       &       \\ 
        &       &       &       &    99 &   227 &   385 &   460 &   339 &   224 &    95 &    54 &       &       &       &       &       \vspace{2pt}\\ 
$-$1.15 &       &       &  8.77 &  8.82 &  8.90 &  8.95 &  8.99 &  9.03 &  9.05 &  9.07 &  9.07 &  9.09 &       &       &       &       \\ 
        &       &       &  0.07 &  0.08 &  0.07 &  0.07 &  0.07 &  0.06 &  0.06 &  0.07 &  0.08 &  0.07 &       &       &       &       \\ 
        &       &       &    58 &   159 &   389 &   694 &   881 &   818 &   672 &   414 &   214 &    73 &       &       &       &       \vspace{2pt}\\ 
$-$1.00 &       &       &  8.76 &  8.82 &  8.88 &  8.94 &  8.98 &  9.02 &  9.04 &  9.05 &  9.06 &  9.07 &  9.09 &       &       &       \\ 
        &       &       &  0.11 &  0.07 &  0.07 &  0.07 &  0.06 &  0.06 &  0.06 &  0.06 &  0.07 &  0.07 &  0.07 &       &       &       \\ 
        &       &       &    76 &   233 &   602 &   958 &  1226 &  1455 &  1331 &   922 &   470 &   224 &    80 &       &       &       \vspace{2pt}\\ 
$-$0.85 &       &       &  8.73 &  8.81 &  8.88 &  8.94 &  8.98 &  9.02 &  9.04 &  9.05 &  9.06 &  9.06 &  9.08 &       &       &       \\ 
        &       &       &  0.11 &  0.08 &  0.08 &  0.07 &  0.06 &  0.06 &  0.06 &  0.06 &  0.06 &  0.06 &  0.07 &       &       &       \\ 
        &       &       &   120 &   316 &   662 &  1152 &  1639 &  1898 &  1996 &  1514 &   966 &   467 &   162 &       &       &       \vspace{2pt}\\ 
$-$0.70 &       &  8.70 &  8.75 &  8.79 &  8.87 &  8.93 &  8.98 &  9.02 &  9.03 &  9.05 &  9.06 &  9.07 &  9.08 &  9.09 &       &       \\ 
        &       &  0.13 &  0.10 &  0.08 &  0.08 &  0.08 &  0.06 &  0.06 &  0.06 &  0.06 &  0.06 &  0.06 &  0.06 &  0.06 &       &       \\ 
        &       &    71 &   165 &   341 &   705 &  1223 &  1809 &  2214 &  2550 &  2338 &  1650 &   913 &   334 &   128 &       &       \vspace{2pt}\\ 
$-$0.55 &       &  8.70 &  8.76 &  8.79 &  8.86 &  8.93 &  8.98 &  9.01 &  9.03 &  9.05 &  9.06 &  9.06 &  9.07 &  9.07 &  9.09 &       \\ 
        &       &  0.10 &  0.11 &  0.10 &  0.08 &  0.07 &  0.07 &  0.06 &  0.05 &  0.05 &  0.05 &  0.06 &  0.06 &  0.06 &  0.06 &       \\ 
        &       &    82 &   195 &   368 &   654 &  1130 &  1703 &  2296 &  2743 &  2955 &  2503 &  1671 &   815 &   271 &    84 &       \vspace{2pt}\\ 
$-$0.40 &  8.56 &  8.65 &  8.71 &  8.77 &  8.85 &  8.92 &  8.97 &  9.01 &  9.03 &  9.05 &  9.06 &  9.06 &  9.07 &  9.07 &  9.08 &       \\ 
        &  0.12 &  0.10 &  0.11 &  0.10 &  0.09 &  0.08 &  0.07 &  0.06 &  0.06 &  0.05 &  0.05 &  0.05 &  0.05 &  0.06 &  0.06 &       \\ 
        &    63 &   123 &   182 &   367 &   628 &  1010 &  1509 &  2100 &  2826 &  3186 &  3005 &  2197 &  1257 &   497 &   148 &       \vspace{2pt}\\ 
$-$0.25 &       &  8.58 &  8.69 &  8.78 &  8.83 &  8.90 &  8.96 &  9.01 &  9.03 &  9.05 &  9.06 &  9.06 &  9.07 &  9.07 &  9.08 &  9.06 \\ 
        &       &  0.12 &  0.12 &  0.10 &  0.10 &  0.09 &  0.08 &  0.07 &  0.05 &  0.05 &  0.05 &  0.05 &  0.05 &  0.05 &  0.06 &  0.05 \\ 
        &       &   108 &   177 &   324 &   550 &   734 &  1113 &  1673 &  2327 &  2826 &  2830 &  2466 &  1598 &   721 &   236 &    57 \vspace{2pt}\\ 
$-$0.10 &       &  8.56 &  8.64 &  8.74 &  8.81 &  8.88 &  8.94 &  8.99 &  9.03 &  9.05 &  9.06 &  9.06 &  9.07 &  9.07 &  9.07 &  9.07 \\ 
        &       &  0.09 &  0.13 &  0.11 &  0.09 &  0.09 &  0.08 &  0.07 &  0.06 &  0.05 &  0.05 &  0.05 &  0.05 &  0.05 &  0.06 &  0.05 \\ 
        &       &    76 &   115 &   284 &   405 &   580 &   837 &  1335 &  1693 &  2204 &  2464 &  2246 &  1594 &   803 &   293 &    61 \vspace{2pt}\\ 
$+$0.05 &       &  8.51 &  8.58 &  8.69 &  8.79 &  8.85 &  8.92 &  8.97 &  9.02 &  9.04 &  9.06 &  9.07 &  9.07 &  9.07 &  9.07 &  9.07 \\ 
        &       &  0.12 &  0.13 &  0.10 &  0.11 &  0.09 &  0.08 &  0.07 &  0.06 &  0.05 &  0.05 &  0.05 &  0.05 &  0.05 &  0.06 &  0.06 \\ 
        &       &    49 &    98 &   151 &   287 &   416 &   591 &   816 &  1178 &  1446 &  1831 &  1760 &  1388 &   810 &   332 &   103 \vspace{2pt}\\ 
$+$0.20 &       &       &  8.53 &  8.66 &  8.72 &  8.82 &  8.89 &  8.96 &  8.99 &  9.04 &  9.05 &  9.07 &  9.07 &  9.06 &  9.07 &  9.06 \\ 
        &       &       &  0.12 &  0.12 &  0.10 &  0.09 &  0.09 &  0.08 &  0.07 &  0.05 &  0.05 &  0.05 &  0.04 &  0.05 &  0.05 &  0.06 \\ 
        &       &       &    63 &   110 &   179 &   327 &   384 &   530 &   740 &   913 &  1104 &  1179 &   944 &   622 &   307 &   103 \vspace{2pt}\\ 
$+$0.35 &       &       &       &  8.59 &  8.69 &  8.77 &  8.85 &  8.92 &  8.98 &  9.01 &  9.04 &  9.06 &  9.07 &  9.07 &  9.07 &  9.06 \\ 
        &       &       &       &  0.11 &  0.10 &  0.10 &  0.08 &  0.07 &  0.07 &  0.06 &  0.05 &  0.04 &  0.04 &  0.05 &  0.05 &  0.05 \\ 
        &       &       &       &    65 &   116 &   224 &   296 &   360 &   385 &   559 &   703 &   708 &   658 &   428 &   200 &    66 \vspace{2pt}\\ 
$+$0.50 &       &       &       &       &  8.64 &  8.72 &  8.82 &  8.88 &  8.94 &  9.00 &  9.03 &  9.05 &  9.07 &  9.07 &  9.08 &       \\ 
        &       &       &       &       &  0.12 &  0.11 &  0.09 &  0.08 &  0.07 &  0.06 &  0.05 &  0.05 &  0.04 &  0.04 &  0.04 &       \\ 
        &       &       &       &       &    92 &   137 &   202 &   236 &   267 &   308 &   368 &   384 &   356 &   232 &    98 &       \vspace{2pt}\\ 
$+$0.65 &       &       &       &       &  8.63 &  8.71 &  8.79 &  8.86 &  8.90 &  8.97 &  9.00 &  9.04 &  9.06 &  9.07 &  9.05 &       \\ 
        &       &       &       &       &  0.12 &  0.12 &  0.08 &  0.07 &  0.08 &  0.07 &  0.06 &  0.05 &  0.04 &  0.04 &  0.04 &       \\ 
        &       &       &       &       &    55 &    77 &    98 &   146 &   136 &   158 &   162 &   187 &   154 &   131 &    62 &       \vspace{2pt}\\ 
$+$0.80 &       &       &       &       &       &       &       &  8.84 &       &  8.93 &  8.98 &  9.02 &  9.04 &       &       &       \\ 
        &       &       &       &       &       &       &       &  0.10 &       &  0.07 &  0.06 &  0.06 &  0.04 &       &       &       \\ 
        &       &       &       &       &       &       &       &    63 &       &    85 &    68 &    68 &    72 &       &       &       \\ 

\hline
\end{tabular}
\normalsize
\end{table*}
%--------------------------------------------------------------

The final scatter is consistent
with the intrinsic uncertainties in the measure of metallicity 
($\sim$0.03~dex for the calibration, to be added to the uncertainties 
in the line ratios), on mass (estimated to be 0.09~dex by \citealt{Tremonti04}),
and on the SFR, which are dominated by the uncertainties on dust extinction.
Nevertheless, the scatter around FMR tends to reduce when the minimum 
redshift z$_{min}$ of the galaxy sample is increased.
This means that part of the residual scatter is probably due to the 
different apertures used to measure mass, based on a total magnitude,  
and metallicity and SFR, derived for the central 3\arcsec. 
In particular, the effect of metallicity gradients are expected to 
become less important at larger redshifts, and for this reason the increase 
of z$_{min}$ is able to reduce the scatter.

A close inspection of fig.~\ref{fig:plotdif} reveals the presence of 
an extended wing
toward lower metallicities in the distribution of scatter. This extension 
contains $\sim$3\% of the objects. Most of these galaxies 
have low \mstar and high SSFR, and could be objects in special conditions. 
For example, 
they could be interacting galaxies, which will be discussed in 
sec.\ref{sec:discussion}.\\

We have fit the median values of metallicity of the SDSS galaxies 
in table~1 with a second-order polynomial in \mstar\ and
SFR, obtaining:
\begin{equation}
\label{eq:fit}
\begin{array}{rl}
12+log(O/H)=&8.90+0.37m-0.14s-0.19m^2\\
            &+0.12ms-0.054s^2\\
\end{array}
\end{equation}
where $m$=log(\mstar)--10 and $s$=log(SFR) in solar units.
The residual scatter of median metallicities around this fit is 0.001~dex.
Such a fit, shown in fig.~\ref{fig:cfr1},
provides a clear representation of the dependence of metallicity both on 
\mstar\ and SFR, and allows to compare the local FMR with high redshift galaxies.\\

The shape of the FMR surface depends on a number of factors, such as 
the selection of the
galaxy sample and the way metallicity, \mstar\ and SFR are measured.
We have done a number of checks to test whether the result depends critically 
on either of our assumptions.  
First, we changed the thresholds in SNR and redshift used to select the galaxy
sample, and checked that the results do not change systematically
with these thresholds.
Second, we have studied the effect of considering only metallicities 
derived either from R23 
or from \nii/\ha.  As discussed in sec.~\ref{sec:sdss}, systematic differences 
are found but are limited to the level of 0.05~dex. 
Apart from this metallicity offset,
the shape of the FMR does not change by more than 0.05~dex at any point.
In particular, there is no large, monotonic dependence of the difference
with SFR or SSFR. This is interesting because there could be systematic 
effects related either to the density of the HII regions or to the 
ionization parameter U, which could depend 
on the SSFR (e.g., \citealt{Shapley05a,Hainline09})
and introduce spurious behavior. R23 and \nii/\ha\ show opposite 
dependencies with U (see, for example, \citealt{Liu08} and \citealt{Nagao06}). 
For this reason the absence of 
systematic differences between these two line ratios with SSFR is an indication 
that the
dependence of measured metallicity on density or U, if present, is small, 
in agreement with
the findings of \cite{Liu08} and \cite{Brinchmann08}.\\

There are two points that could affect the shape of the FMR. 
First, SFR is estimated from \ha\ luminosity corrected for extinction using 
the Balmer decrement.
Several authors \citep{Kennicutt98,Moustakas06} have shown that this is 
a reliable SFR indicators over a large range of galaxy properties. Others
\citep{Charlot01,Brinchmann04} have discussed that systematic effects with mass 
and metallicity could be present. As we have split galaxies in bins of mass 
and SFR, and inside each bin metallicity spans a small range, any 
systematic effect on SFR does not hamper the existence of the FMR but 
could change its shape. 
Second, we are using 
SFRs and metallicities that apply only to the 
central 3\arcsec of the galaxies, corresponding to 4--11 kpc projected angular 
size given our redshift range.
These quantities are compared with total mass derived from integrated
photometry
\footnote{see http://www.mpa-garching.mpg.de/SDSS/DR7/masscomp.html} 
\citep{Kauffmann03a,Salim07}. 
In the SDSS sample, the fraction of mass contained inside the projected fiber 
aperture (as listed in the MPE/JHU catalog) is about 1/3 of the total, with 
log(total mass) -- log(fiber mass) = 0.50$\pm$0.15,
and no systematic dependence on mass.
We find that total and fiber mass are correlated with SFR and metallicity 
at a similar degree, and we used the total mass as this is the quantity
usually available in galaxy catalogs.
The measured metallicity is expected to be a fair representation of total metallicity:
abundance gradients can be present but, at least in local galaxies, they
are of modest importance in this contest, both because only weak variations 
with radius are usually found (e.g., \citealt{Magrini09})
and because our apertures sample a significant 
part of the galaxies. Correcting fiber SFR into total ones is not straightforward, as
there is no guarantee that \ha\ scales radially as luminosity or mass.
If SFR scales with radius as mass, we expected that 
total SFR are 3 times larger than the fiber ones considered here.
The use of total SFR would produce a FMR shifted towards higher SFRs, 
and its shape could have some changes.
Nevertheless the small scatter observed in the FMR means that the fiber 
SFR must be well correlated with the total one. 

Summarizing, even if the overall shape of the FMR can change in 
different samples of galaxies and depends on several details, 
the main properties of the FMR are 
very robust and passed all our tests.

%---------------------------------------------------------------
\begin{figure*}
\centerline{
	\includegraphics[width=0.48\textwidth]{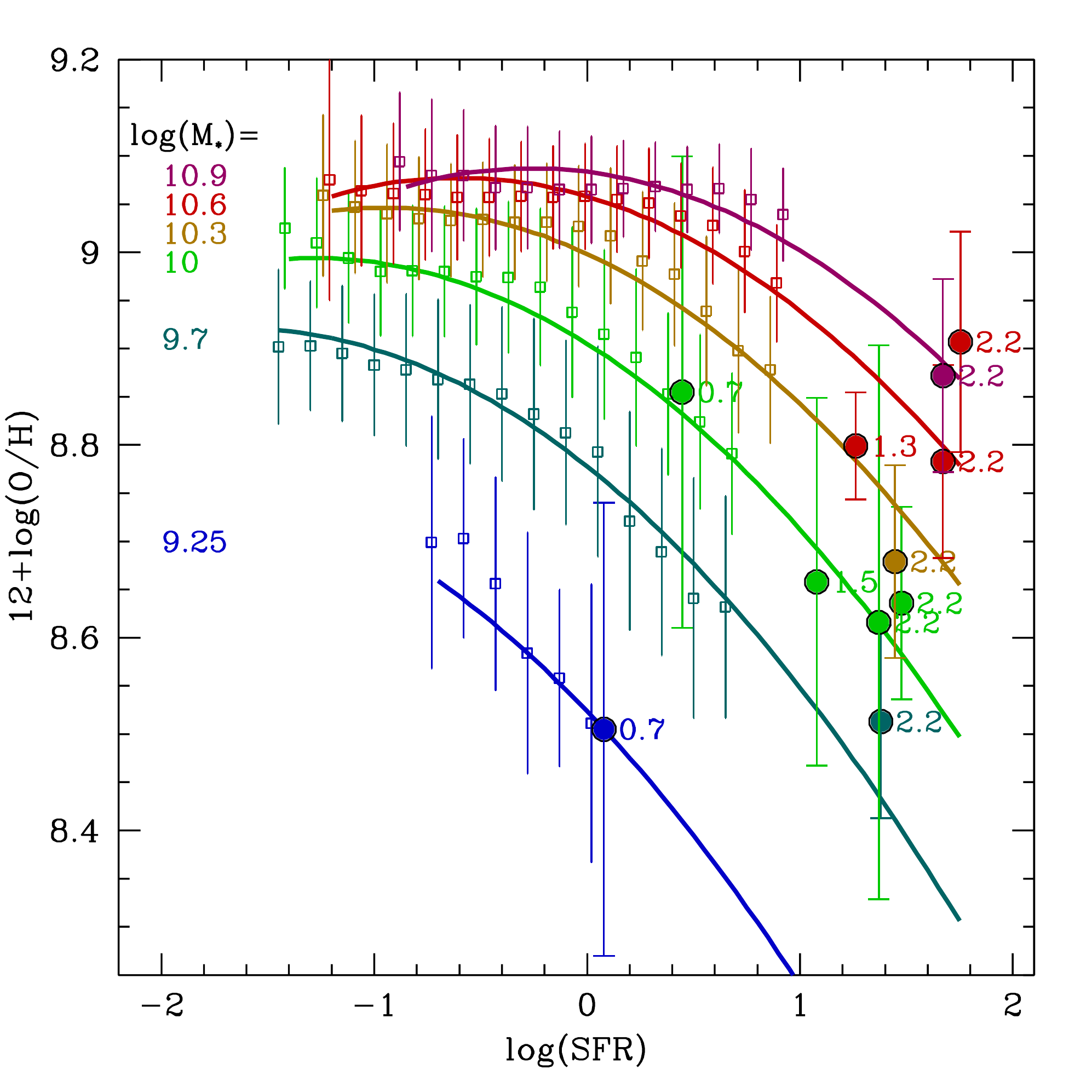} 
	\includegraphics[width=0.48\textwidth]{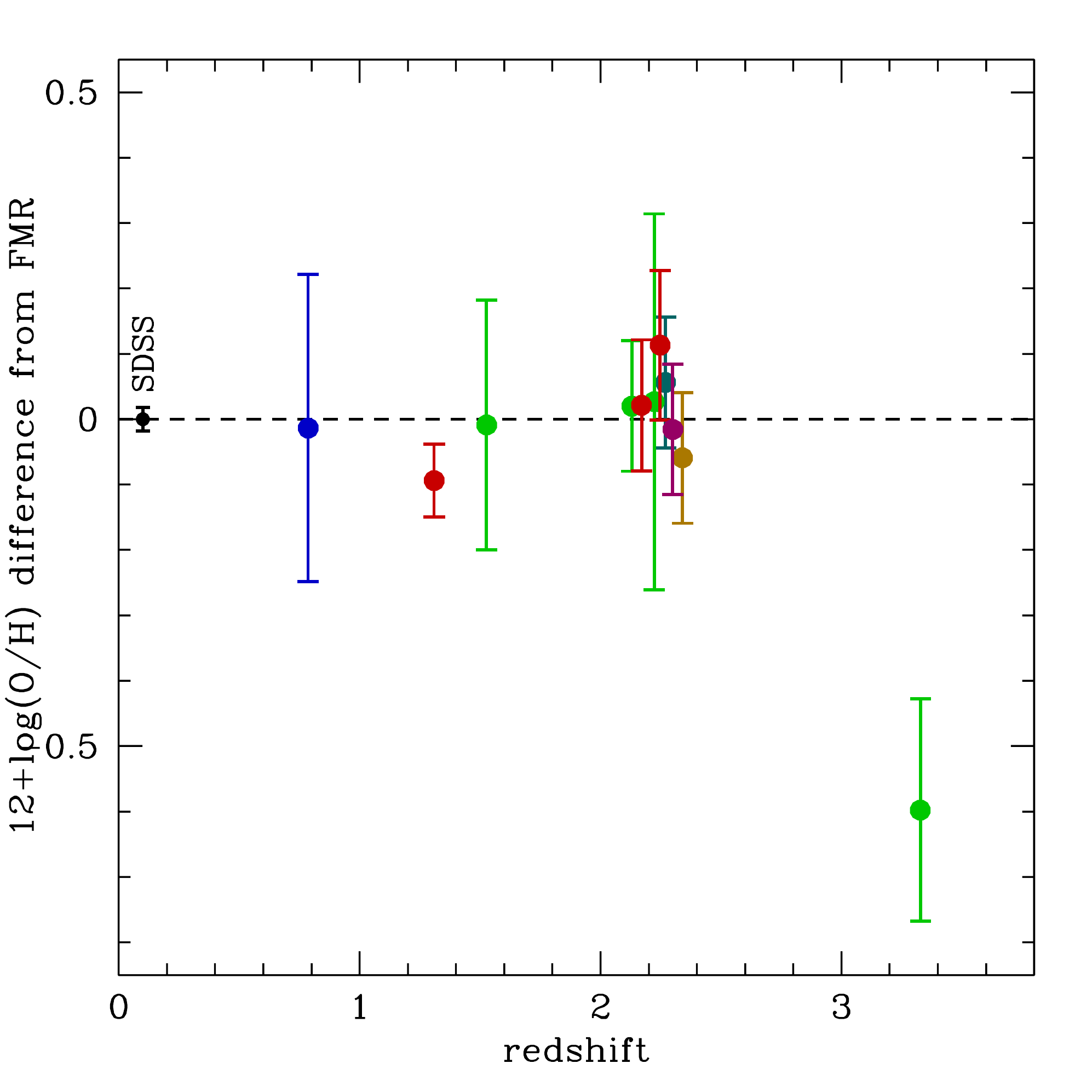} 
}
\caption{
{\em Left:} Metallicity as a function of SFR for galaxies in the three bins 
of \mstar\ containing high-redshift galaxies. The values of log(\mstar) 
are shown by the labels on the left.
Empty square dots are the median values of metallicity of local SDSS galaxies, 
with error bars showing 1$\sigma$ dispersions. 
Lines are the fits to these data.
Solid dots are median values for high-redshift galaxies with z$<$2.5 in the 
same mass bins, with labels showing redshifts.
{\em Right:}
metallicity difference from the FMR for galaxies at different redshifts,
color-coded in mass as in the left panel.
The SDSS galaxies defining the relation are showing at z$\sim$0.1 with their
dispersion around the FMR. All the galaxy samples up to z=2.5 are consistent 
with no evolution of the FMR defined locally. 
Metallicities lower by $\sim$0.6~dex are observed at z$\sim$3.3.
}
\label{fig:plotevol}
\end{figure*}

%====================================================================
\section{The local FMR and high-redshift galaxies}
\label{sec:highz}

Using the samples described in Sect.~\ref{sec:z2} and \ref{sec:z3}
we compare the dependence of metallicity on \mstar\ and SFR in 
galaxies at z$\sim$0.8, z$\sim$1.4, z$\sim$2.2 and z$\sim$3.3,
and compare it with the properties of local SDSS galaxies. 
Galaxies at all redshifts follow well defined mass-metallicity relations
(see, for example, \citealt{Mannucci09b}, and references therein).
For this reason each of these samples, 
except the one an z$\sim$3.3 that contains 16 objects only,
is divided into two equally-numerous samples
of low- and high-\mstar\ objects. Median values of \mstar, SFR and 
metallicities are computed for each of these samples.

Galaxies at intermediate and high redshifts 
show, on average, larger SFR with respect to local SDSS galaxies. 
This is easily explained by selection, because 
only galaxies with significant SFRs are selected and observed spectroscopically.
Galaxies at z$\sim$0.8 have values of \mstar\ and SFR which overlap with the SDSS 
sample, and therefore the two samples can be directly compared.
These galaxies are found to be completely consistent with the FMR defined 
by SDSS galaxies, with no evidence for evolution.
This is shown in  Figs.~\ref{fig:cfr1} and \ref{fig:plotevol}.
At redshift above 1, some extrapolation towards higher SFRs
of the fit in eq.~\ref{eq:fit} in required. All galaxies 
are within 0.6~dex from the most active SDSS galaxies, while 
most massive galaxies at z=2.2 require an extrapolation of 1~dex. 
For comparison, SDSS galaxies span two orders-of-magnitude in SFR
(see Fig.\ref{fig:plotevol}).
The necessity of an extrapolation introduces some uncertainty, but we have checked
the the result does not depend critically on the characteristic of the fit, 
such as the degree of the used polynomial. 

When taking into account the uncertainties, data up to z$\sim$2.5 are 
consistent, both in shape and in normalization, 
with the same FMR defined by SDSS, with no evidence for evolution, 
Distant galaxies show larger dispersions
than the local SDSS galaxies, between 0.2 and 0.3~dex.
At least part of these relatively larger dispersions are due to the large 
uncertainties in the estimates of both metallicity and SFR, 
but part of it could be intrinsic, related to different
evolutionary stages of the galaxies. 
A larger sample of well measured galaxies is needed to address this point.\\

The existence of a relation between mass, SFR and metallicity
could be considered a mere consequence 
of the existence of two other well-known relations, the mass-metallicity 
relation, and the mass-SFR relation
(see, for example, \citealt{Schiminovich07}). In fact, the novelty of 
this relation 
can be understood by comparing the relative redshift evolutions:
while both the mass-metallicity and the mass-SFR relations are known to evolve 
significantly with redshift 
\citep{Daddi07,Mannucci09b,Forster-Schreiber09}, the  FMR
remains the same up to z=2.5, a period of time spanning 80\% of the 
universe lifetime.
Under this respect, the FMR seems to be the fundamental one,
directly related to the mechanisms of galaxy formation.\\

In the SDSS sample, metallicity changes more with \mstar\ ($\sim$0.5~dex 
from one extreme to the other at constant SFR, see Fig.~\ref{fig:massmet})
than with SFR ($\sim$0.30~dex at constant mass). 
Therefore mass is the main driver of the level of chemical enrichment 
of SDSS galaxies.
This is related to the fact that 
galaxies with high SFRs, the objects showing the strongest dependence 
of metallicity on SFR (see the right panel of fig.~\ref{fig:massmet}), 
are quite rare in the local universe.
At high redshifts, mainly active galaxies are selected and the dependence
of metallicity on SFR becomes dominant. 

%---------------------------------------------------------------------------------
\subsection{Possible effects on data at z$>$2.5} 
\label{sec:evol}

Galaxies at z$\sim$3.3 show metallicities 
lower of about 0.6~dex with respect to
both the FMR defined by the SDSS sample and galaxies at 0.5$<$z$<$2.5. 
This is an indication that some evolution of the FMR 
appears at z$>$2.5, although its size 
can be affected several potential biases of different nature 
that should be taken into careful account.

First, metallicity at z=3.3 is measured by the oxygen indicators only, 
while SDSS galaxies use both R23 and \nii/\ha. 
As discussed in sec.~\ref{sec:z3}, in the local universe
both indicators give consistent results, with systematic differences limited 
to 0.05~dex.
Also, galaxies at z$\sim$0.8 use the same metallicity based on R23.
Therefore this is not likely to be the origin of the evolution of 0.6~dex,
although it could be responsible for part of the difference.

Second, systematic evolution with redshift of the photoionization conditions
could be present, for example because high-redshift galaxies have larger 
SFRs. The presence of such an effect can be studied by
comparing different line ratios, and several
studies up to z$\sim$2.5 indicate that such an evolution actually
exists \citep{Shapley05a,Erb06a,Brinchmann08,Liu08,Hainline09} .
Nevertheless, this effect is not large enough to move galaxies
up to z=2.2 off the FMR, even if some of these galaxies have SFR larger than 
most of the galaxies at z=3.3. This is in agreement with the results by 
\cite{Liu08,Brinchmann08}
who estimate a minor effect of this evolution on the measure of metallicity.
Therefore evolution of the conditions
could give some contribution to the observed evolution but are not likely 
to be the only reason.

Third, at z=3.3 the \ha\ and \nii\ lines fall at $\sim$2.8\mic\ 
and are not observed 
in any galaxy of this group. For this reason we cannot use this line ratio to remove 
AGNs. X-ray and mid-IR data on these targets have been analyzed in order to exclude 
dominant AGN \citep{Maiolino08}, but it is possible that some faint AGN is still present
among these galaxies \citep{Wright10}. The presence of such objects would tend to reduce the
measured metallicity. As almost all galaxies at z$\sim$3 have metallicity 
below both the FMR
and the mean values at z$\sim$2 \citep{Mannucci09b}, 
this effect can explain the 
observed difference only if AGNs are present in most galaxies.

%---------------------------------------------------------------
\begin{figure*}
\centerline{
	\includegraphics[width=0.48\textwidth]{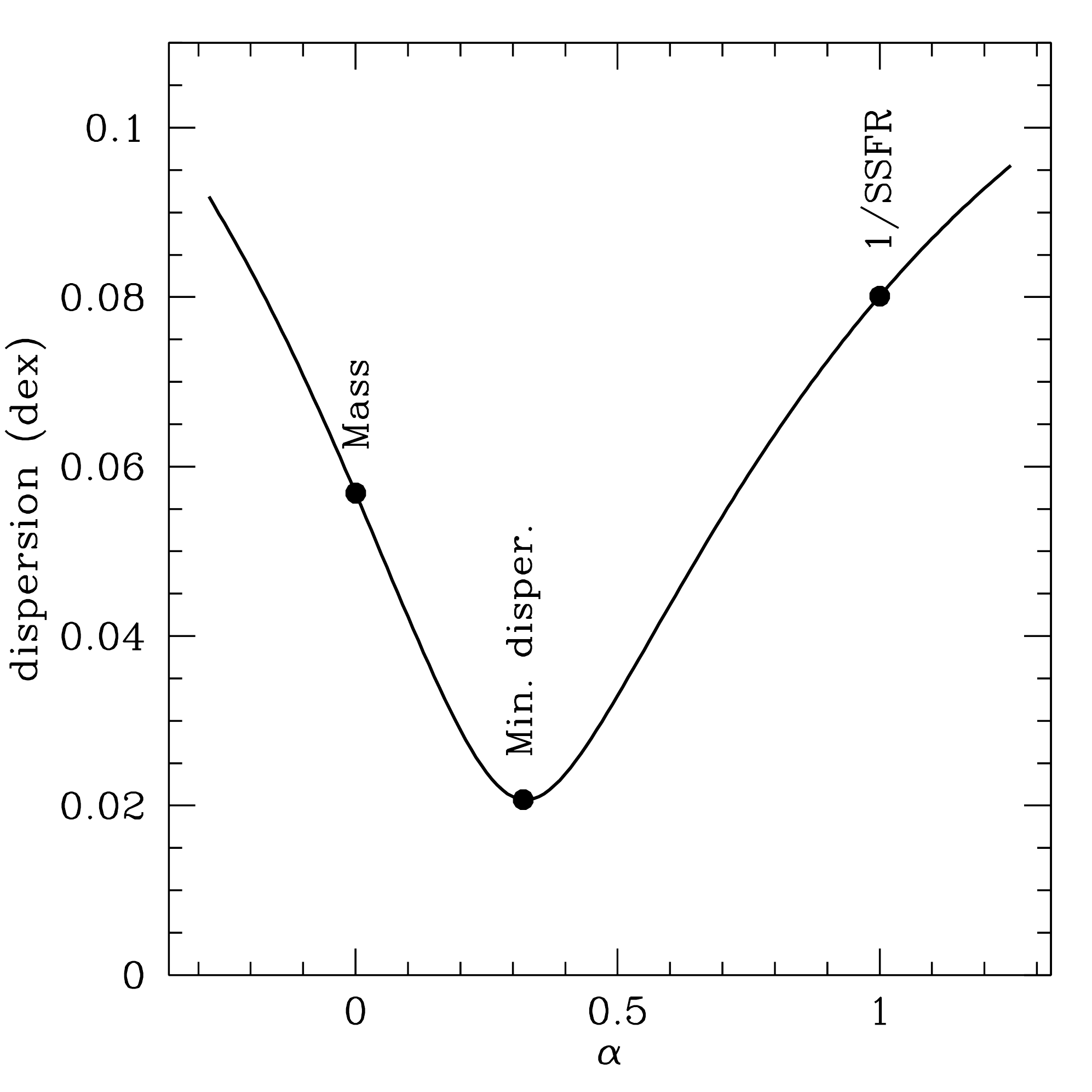}
	\includegraphics[width=0.48\textwidth]{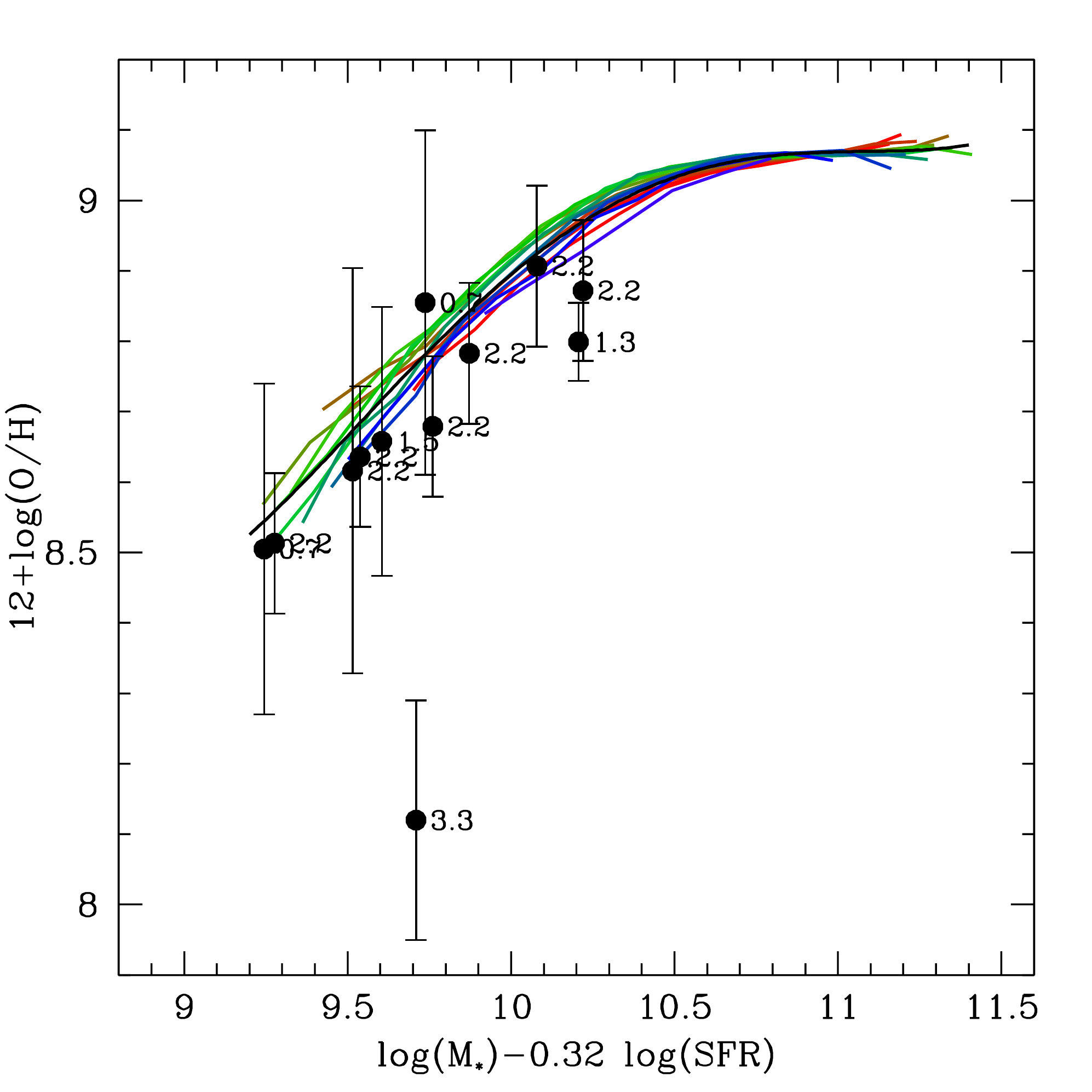}
}
\caption{
{\em Left:}
Residual dispersion of the median values of metallicity of SDSS galaxies 
as a function of $\alpha$ as defined in eq.~\ref{eq:def}. 
The values corresponding to the minimum dispersion ($\alpha$=0.32),
to $\alpha$=0  ($\mu_\alpha$=log(\mstar)) and 
and $\alpha$=1 ($\mu_\alpha$=--log(SSFR)) are shown.
{\em Right:} metallicity as a function of $\mu_{0.32}$, which minimize the 
residual scatter. Colored lines are local SDSS galaxies, with colors as in the left panel 
of Fig.~\ref{fig:massmet}. The black line shows the polynomial best fit. 
Black dots are high-redshift galaxies, labelled with redshifts. These galaxies, except
those at z$\sim$3.3, appear to follow the same relation define by SDSS galaxies.
}
\label{fig:bestfit}
\end{figure*}
%------------------------------------------------------------------------

Fourth, 
observations at z$\sim$3.3 target the \oiiib\ line, which 
has a strong dependence on metallicity, with more metal-poor regions emitting
a brighter \oiiib\ line for a given SFR \citep{Maiolino08}. 
As only the brightest part of the galaxies are detected in our data, this effect, 
could bias the measured line ratios 
towards the regions of lower metallicities.
At lower redshift this effect could be less important, both because
\oiiib\ is usually intrinsically fainter than \ha\, observed at 
z$<$2, and because the cosmological dimming of the surface brightness, 
proportional to $(1+z)^4$, is a much more severe problem at higher redshift.
In our data at z=3.3, metallicity does not seem to increase with 
increasing photometric aperture, although we are limited by the low SNR
in the external regions of the galaxies.
This effect could be present but is not likely to produce the 
observed difference of a factor of 3.

Finally, there could be selection effects resulting in a reduction of
the average metallicity of observed sample. For example,
if more metal-rich galaxies also have larger amounts of dust, 
it is possible that our UV-selected galaxies at
z$\sim$3.3 preferentially select low-metallicity objects with lower
dust column densities. If present, this effect would also work at lower redshift, 
introducing some evolution also at z=2.2. Again, such an effect could be present but
is unlikely to produce a strong differential evolution between z=2.2 and z=3.3.\\

In conclusion, even if the size of the evolution at z=3.3 could be affected by 
several problems, it is unlikely to be totally due to observational biases. \\

%===========================================================================

\section{A projection of the FMR that remove secondary dependencies}
\label{sec:project}

At a given mass, galaxies with higher SFR have lower metallicities and,
therefore, have the metallicity properties of lower mass galaxies.
As a consequence, we expect that a combination of \mstar\ and SFR could be 
better correlated with metallicity.
This can be seen in the central panel of fig.~\ref{fig:cfr1}, showing a
projection of the FMR which considerably reduces the metallicity scatter.
To  investigate this point we introduce a new
quantity $\mu_\alpha$ 
obtained as linear combination of SFR and \mstar\  as:
\begin{equation}
\mu_\alpha =\rm{log}(M_*)-\alpha~\rm{log(SFR)}
\label{eq:def}
\end{equation}
where $\alpha$ is a free parameter. For $\alpha$=0, $\mu_0$ corresponds to
log(\mstar), while for $\alpha$=1, $\mu_1$=--log(SSFR).

The value of $\alpha$ that minimizes the scatter of median metallicities of 
SDSS galaxies around the relation corresponds to the 
quantity $\mu_\alpha$ that is more directly correlated with metallicity.
Fig.~\ref{fig:bestfit} shows the scatter of data in table~1
as a function of $\alpha$. 
These results show that neither \mstar\  ($\alpha$=0),
nor, SSFR ($\alpha$=1) are the quantities producing the smallest scatter. 
In fact, $\alpha\sim0.32$ produces a minimum in the dispersion.
The resulting diagram is shown in the right panel of Fig.~\ref{fig:bestfit}, 
where metallicity is plotted against $\mu_{0.32}$.
The median of the resulting distribution cab be fitted by:
\begin{equation}
\label{eq:cfr2}
12+log(O/H) =  8.90  + 0.39x -0.20x^2 - 0.077x^3 +0.064x^4
\end{equation}
where $x=\mu_{0.32}-10$.

Even if the minimization is computed with SDSS galaxies only, 
it turns out that high-redshift galaxies up to z=2.5 follow 
the same relation between $\mu_{0.32}$ and metallicity as in the local universe,
and also have the same range of values of $\mu_{0.32}$. 
This is the same effect noted in the previous section, where high-redshift galaxies have 
been found to follow the extrapolation of the FMR, 
but with two important changes: first, no extrapolation 
from the SDSS galaxies is now needed, because both samples have similar values of $\mu_{0.32}$; 
second, it is possible to search for simple physical interpretation
of $\mu_{0.32}$ in terms of the physical processes in place.

In practice, metallicity of star-forming galaxies of any mass, any SFR and at any redshift 
up to z=2.5 follow the following relation:
\begin{equation}
\label{eq:cfr2}
\begin{array}{rll}
12+log(O/H)=&8.90+0.47x  & ~\rm{if}~~ \mu_{0.32}<10.2 \\
            &9.07        & ~\rm{if}~~ \mu_{0.32}>10.5 \\
\end{array}
\end{equation}
with $x=\mu_{0.32}-10$.

%---------------------------------------------------------------------------------
\subsection{Metallicity and SSFR}

\begin{figure}
\centerline{
   \includegraphics[width=\columnwidth]{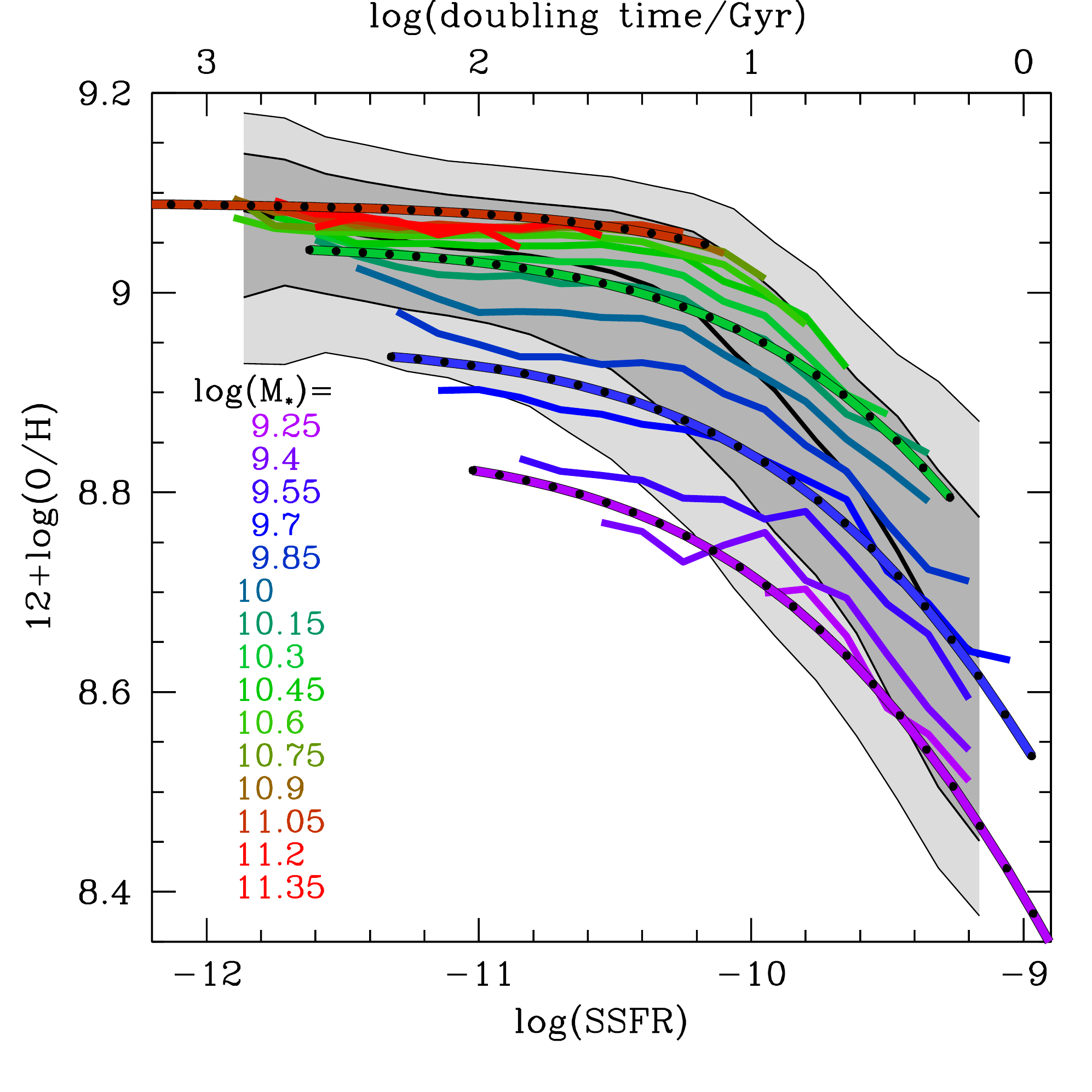} 
}
\caption{
Metallicity of SDSS galaxies as a function of SSFR.
The grey-shaded areas contain 64\% and 90\% of all SDSS galaxies, with the 
thick central line showing the median relation. The colored lines show the 
median metallicities for different values of \mstar.
The lines with black dots show fits to these metallicity distribution for four values of 
log(\mstar), 9.4, 9.7, 10, and 10.9. The model is described in sec.~\ref{sec:infall}
and includes only dilution for infalling gas and the SK law.
}
\label{fig:ssfrmet}
\end{figure}

It is interesting to plot metallicity as a function of SSFR, as in
fig.~\ref{fig:ssfrmet}, because 
several properties of galaxies depend on this quantity.
In this plot it can be seen that SDSS galaxies of any mass have the
same dependence of metallicity on SSFR. A threshold SSFR exists, about 
$10^{-10}$yr$^{-1}$ which discriminates the abundance effect of the SFR.
Above this limit, metallicity decreases rapidly with SFR in galaxies on any mass.
Below this limit, metallicity is constant in massive galaxies (log(\mstar)$>$10.5) 
and slowly decreasing with SFR in less massive galaxies. 
The interpretation of this behavior is given in the next section.

The fraction of galaxies above and below this SSFR threshold changes with \mstar.
Most of the low-mass galaxies, and only a small fraction of high-mass galaxies 
in our SDSS sample are in the ``high SSFR'' regime, and this is the well-known 
"downsizing" effect
\citep{Gavazzi96,Cowie96}. 
As \mstar\ and SFR are independent variables of the FMR, downsizing does not 
shape the relation but defines how it is populated, i.e, how many galaxies 
of a given \mstar\ 
have a certain level of SFR and, as a consequence, metallicity.

%====================================================================================
\section{Discussion}
\label{sec:discussion}

In sec.~\ref{sec:cfr} we have shown that in the local universe a tight relation exists 
between metallicity, stellar
mass, and SFR, in which metallicity increases with \mstar\ and decreases with SFR
in a systematic way. In sec.~\ref{sec:highz} we have shown that the same relation, 
without any evolution, holds up to z=2.5, and that the observed evolution of the 
mass-metallicity relation is simply due to the sampling of different parts of this 
relation at different redshifts. In sec.~\ref{sec:project} we have seen that 
the systematic dependence of metallicity on \mstar\ and SFR at all redshifts
can be expressed in an easy form by introducing the quantity $\mu_{0.32}$, linear 
combination of \mstar\ and SFR. 

The interpretation of these results must take into account several effects.
In principle, metallicity is a simple quantity as it is dominated by three processes:
star formation, infall, outflow. If the scaling laws of each of these three processes are
known, the dependence of metallicity on SFR and \mstar\ can be predicted.
In practice, these three processes have a very complex dependence of the properties of the 
galaxies, and can introduce scaling relations in many different ways. 

First, it is not known how {\em outflows} depend on the properties of the galaxies.
In many models, star formation produces SNe which inject energy, radiation and momentum 
into the interstellar medium, with the result of ejecting part of the enriched gas
\citep{Veilleux05,Spitoni10}.
A central Active Galactic Nucleus (AGN) can also provide feedback \citep{Somerville08}.
The properties of this galactic winds are debated. \cite{Dekel03} reproduced the 
mass-metallicity relation with a wind related to the energy of SNe, which is proportional 
to stellar mass. In large galaxies with deep potential wells, such a wind is not effective
in producing an outflow \citep{Tremonti04}. 
\cite{Murray05}, \cite{Dave07a} and \cite{Oppenheimer08} 
discuss a different scheme, the ''momentum-driven wind'',
in which wind speed increases with galaxy mass while its efficiency decreases.

Second, {\em infalls} are expected to influence metallicity in two ways. On 
the one hand, infall of metal poor gas directly reduces the observed metallicity by
diluting the metal-rich gas, as discussed, for example, by 
\cite{Finlator08} in the context of their wind scheme.
On the other hand, infall is expected to produce star-formation activity following the 
Schmidt-Kennicutt law, and  the metals produced by new stars are expected to increase metallicity. 
If merging between galaxies, rather than smooth infall from the IGM, is the main channel
to drive gas into galaxies, the fuel of star formation can be significantly enriched, 
and the dilution effect could be absent.

%---------------------------------------------------------------
%-----------------------------------------------------------------

Third, in some semianalytical models 
of galaxy formation \citep{deRossi07,Mouchine08} the behavior of metallicity 
is dominated  by
the dependence of {\em star-formation efficiency} of galaxy mass: 
less massive galaxies are less evolved 
and, therefore, show lower metallicities. Some other models 
\citep{Brooks07,Dayal09,Salvaterra09} put together downsizing and feedback: 
metallicity is mainly related to different star formation efficiencies
in different galaxies, but the efficiency is regulated by SN feedback.
\cite{Tassis08} agree that low star-formation efficiency in low-mass galaxies 
is the main driver of the mass-metallicity relation, but also 
pointed out that mixing of heavy elements in the outer regions of galaxies
could help hiding a significant fraction of metals.
Also \cite{Koppen07} attribute the different abundances
to different levels of productions of metals, but their model include
systematic variation of IMF, which is proposed to be more top-heavy in galaxies 
with higher SFRs.

A full exploitation of our results requires a full discussion
of these models, that is beyond the scope of this paper and will be subject of a future work. 
Here we discuss some simple implications of our results on the role and the properties 
of infalls and outflows. In the next two subsections we will assume two limiting cases 
for the timescale of chemical enrichment as compared to the other relevant timescales. 

%----------------------------------------------------------------------------
\subsection{The effect of the infall}
\label{sec:infall}

The dependence of metallicity on SFR can be explained by the dilution effect of the
infalling gas. 
In a very simple model, we assume that galaxies with low SFR have a given
gas fraction $f_g$. When a certain amount $M_{inf}$ of metal-poor gas
is accreted,  galaxies start forming stars at a given SFR defined by the 
Schmidt-Kennicutt (SK) law on the infalling gas.
This is an empirical relation between 
surface densities of star formation and surface density of gas:
\begin{equation} 
\Sigma_{SFR} \sim \Sigma_{gas}^n
\end{equation}
where $n$ has values around 1.4--1.5 both at low and high redshifts
\citep{Kennicutt98,Bouche07,Kennicutt08,Gnedin10a,Verley10}.
For SDSS galaxies, the spectroscopic aperture is the always same and the SK law
becomes a relation between masses, obtaining:
\begin{equation}
\rm{SFR} \sim M_{inf}^n
\end{equation}
In this model, galaxies are observed during the phase when
the dilution effect of the infall is predominant over the metallicity 
enrichment due to new stars, i.e., 
the observed metallicity [12+log(O/H)]$_{obs}$
is related to the initial metallicity [12+log(O/H)]$_{in}$
by the ratio of pre-existing and infalling gas: 
\begin{equation}
[12+\rm{log(O/H)}]_{obs}=[12+\rm{log(O/H)}]_{in}-\rm{log}\left(1+\frac{M_{inf}}{f_g\ M_\star}\right)
\end{equation}
The results of this simple model are presented in fig.~\ref{fig:ssfrmet}.
For simplicity, only 4 values of mass are shown, using 
the best-fitting values of $f_g$=0.3\% and  $M_{inf}$ 
between $10^{5.5}$ and $10^{7.5}$ \msun. This range of values of $M_{inf}$
is the same for all masses and is determined by the range of observed SSFR. 
For each mass, the metallicity level at low SSFR,  [12+log(O/H]$_{in}$,
is a free parameter, whose value is possibly determined by
other effects, such as outflow.
Our very simple model is capable of reproducing all the main properties of 
the FMR:
metallicity reduces with increasing SFR, a threshold of SSFR exists, larger
metallicity effects are produced in smaller galaxies, both above and below the 
threshold, the slope of the relation at high SSFR is exactly as observed. 

For this scenario to work, the timescales of chemical enrichment must be longer than
the dynamical scales of the galaxies, over which the SFR is expected to evolve. 
In other words, galaxies on the FMR are in a {\em transient phase}:
after an infall, galaxies first evolve towards higher SFR 
and lower metallicities. Later, while gas is converted into stars and new metals are produced, 
either galaxies drop out of the sample because they have faint \ha, or evolve toward higher 
values of $\mu_{0.32}$ and higher 
metallicities along the FMR. A detail modelling of this evolution and a comparison of the 
different timescales involved is needed to test if this is a viable explanation.

In this scenario, the dependence of metallicity on SFR is due to infall and dominates at 
high redshifts, where galaxies with massive infalls and large SFRs are found. In contrast,
in the local universe such galaxies are rare, most of the galaxies have low level of accretion,
and abundances are dominated by the
dependence on mass, possibly due to outflow.

%----------------------------------------------------------------------------
\subsection{Properties of outflows}
\label{sec:outflows}

In many local galaxies, timescales of chemical enrichment can be shorter 
than the other relevant timescales (e.g., \citealt{Silk93}), 
and galaxies can be in a {\em quasi steady-state situation},
in which gas infall, star formation and metal ejection occur simultaneously
\citep{Bouche09}.
This is the opposite situation of what is discussed in the previous section.\\

Assuming this quasi steady-state situation, in which infall and SFR are 
slowly evolving with respect to the timescale of chemical enrichment, our results
can be used to derive information on the mechanisms of infall and outflow.
Fig.~\ref{fig:ssfrmet} implies that a process exists that depends 
only on SSFR which is effective in reducing metallicity
from a level that depends on mass.
In a steady-state situation, infall cannot be the only dominant effect.
The exponent $n$ of the SK relation is larger than 1, and this means that the 
efficiency of star formation increases with gas density, i.e., 
more active galaxies should also be more efficient in converting metal-poor gas into stars. 
As a consequence, the SK law alone, applied to the infalling gas, would predict the opposite 
to what is actually observed, i.e., metallicities increasing with SFR at constant mass. 
Some other effect must be present.

One obvious candidate is the presence of outflows, as usually observed in starburst galaxies 
and discussed by many authors. As discussed above, there are several possible
types of galactic winds, which follow different scaling relations with mass   and SFR.
When the dilution by the infalling gas $M_{inf}$ is considered together with enrichment 
due to the SFR, we can reproduce
the dependence of metallicity on $\mu_{0.32}$ by introducing an outflow
proportional to SFR$^s$~\mstar$^{-m}$, with $s$ and $m$ free parameters.
In this case, at the first order we obtain:
\begin{equation}
12+log(O/H) \sim \rm{log}\left(\frac{\rm{SFR}}{M_{inf}\,\rm{SFR}^s\,M_\star^{-m}}\right)
\label{eq:eq5}
\end{equation}

and, using the SK law:
\begin{equation}
12+log(O/H) \sim m\,\rm{log}(M_\star)+(1-\frac{1}{n}-s)\,\rm{log}(\rm{SFR})
\label{eq:eq6}
\end{equation}
where $n=1.5$ is the index of the SK law.
This is the functional form of $\mu_\alpha$, and comparing this equation
with the best-fitting value of $\alpha=0.32$ we obtain $m=1$ and $s=0.65$. In other words,
in a steady-state situation the outflow must to inversely proportional 
to mass and must increase with SFR$^{0.65}$.\\

We note that, using an index $n$ of the SK relation of 1.5, the best-fitting value of $\alpha$=0.32
corresponds almost exactly to $1-1/n=0.33$.
Using this, the simplest way to combine the relevant physical parameters to produce
$\mu_{0.32}$ is:
\begin{equation}
\label{eq:mu032}
\rm{log}\left(\frac{M_{inf}\,M_\star}{SFR}\right) 
           \sim \rm{log}(M_\star)-(1-1/n)\rm{log}(SFR)
           \sim \mu_{0.32}
\end{equation}
In this form, metallicity increases with \mstar, an effect easily attributable either to 
downsizing or to outflow. In contrast, the dependence of metallicity on $M_{inf}/SFR$ is not 
obvious and require complete modelling.

%--------------------------------------------------------------------------------
\subsection{Merging and smooth accretion}
\label{sec:merging}

The small scatter of SDSS galaxies around the FMR derived in sec.~\ref{sec:cfr}
can be used to constrain the characteristics of gas accretion.
For this infall/outflow scenario to work and produce a very small scatter 
round the FMR, two conditions are simultaneously required:
(1) star formation is always associated to the same level of metallicity
dilution due to infall of metal-poor gas; 
(2) there is a relation between the amount of infalling and outflowing gas and 
the level of star formation.
These conditions for the existence of the 
FMR fits into the smooth accretion models proposed by several groups
\citep{Bournaud09,Dekel09}, where continuos infall of pristine gas is the main
driver of the grow of galaxies. In this case, metal-poor gas is continuously accreted
by galaxies and converted in stars, and a long-lasting equilibrium between gas accretion, 
star formation, and metal ejection is expected to established.

In contrast, larger scatter around the FMR are expected in case of merging, 
for two reasons.
First, in interacting systems, part of the gas producing the starburst could be metal-poor 
material due to 
interaction-induced infall \citep{Rupke08,Rupke10},
but part is expected to
be metal-rich material already present inside the interacting galaxies.
In this case the initial dilution is not present, and 
higher metallicities are expected. This is true, in particular, for
the smaller member of a merger between galaxies with very different masses.
The secondary, smaller mass, galaxy is expected to show higher metallicity
because its star formation activity is expected to be fuelled by gas coming from the 
other, larger, more metal-rich, galaxy. 
This is exactly what is observed by 
\cite{Michel-Dansac08}: higher metallicities, above the mass-metallicity relation,
are present in the secondary member of minor mergers.
Second, the level of SFR is related to the properties of the merging galaxies, and 
is expected to vary significantly during the different stages of the interactions.
As a consequence, a large range on SFR is expected for a given level of metallicity
during the merging history of the systems.
Both effects are expected to produce larger spreads
in merging galaxies, and this is what is actually observed
\citep{Kewley06,Rupke08,Michel-Dansac08,Peeples09}. 
As discussed in sec~\ref{sec:cfr}, the presence of interacting galaxies in 
the SDSS could be at the origin of the extended wing in the distribution of
difference with the FMR, and this point will be investigated in a future paper.

The situation at intermediate redshift, up to z=2.5, is less clear. 
Galaxies show larger dispersion in metallicity, but this can be explained by 
larger uncertainties in the measured values of metallicity, mass and SFR.
With the present data sample it is not possible, therefore, to study whether
smooth accretion is dominant up to z=2.5 as in the local universe or if 
merging has a larger impact on dispersion.
 Nevertheless, the absence of any evolution of the FMR up to
these redshift support a single physical process of accretion in all these galaxies. 
At even higher redshift, z$\sim$3.3, when galaxies show large 
scatter {\em and} different average metallicities, it 
is likely that new physical effects become important.

The uncertainties on SFR can be a critical point. Under this respect, the Herschel 
satellite is expected to 
improve significantly the accuracy of the estimate of total
SFRs in a short time. With its contribution it will be possible
to reduce the scatter, especially at high redshifts, study the presence 
of an intrinsic dispersion of properties among the galaxies, 
and obtain a much better characterization 
of the FMR relation at high redshifts.

%------------------------------------------------------------------------
\subsection{Origin of the evolution at z$>$2.5}

The interplay between gas accretion, star formation, and gas outflow seems
to be the same at any redshift up to z=2.5, as all these galaxies 
follow the same FMR. What is important in this context is the ratio 
between the different rates and the various timescale involved: gas infall,
dynamical times, star formation, stellar evolution, supernova explosion, 
chemical mixing, outflow. 
Apparently the relative importance of these processes does not 
evolve up to z=2.5. It is possible that at higher redshifts
this constant balance does not apply any more and 
lower metallicities are observed 
while galaxies evolve towards the FMR.
This could be related to an increasing importance of merging as a way to
drive cold gas into the galaxies, at least in the most luminous objects that
are preferentially selected as LBG. This point can be  addressed
by studying the morphology and the mass-SFR relation in these objects.

It should be noted that the metallicity of galaxies at z=3.3 can also
be reproduced 
by the same model in sec.~\ref{sec:infall}, in which the infall dilution is dominant.
Local galaxies are reproduced by $M_{inf}$ of the order of $10^6-10^7$ \msun, while 
gas masses of the order of $10^9-10^{10}$ \msun\ are needed for galaxies at
z=3.3. This is in very good agreement with what obtained by 
\cite{Mannucci09b} and \cite{Cresci10}. 
Also, galaxies at these high redshifts could be 
preferentially detected during the first stages of the starburst, 
when the dilution effect is maximum.  This is possible
if the starburst has a peak on short timescales, shorter than the timescales 
of metal production and chemical mixing, and declines afterwards. 
Galaxies could also become more dust-rich during the later phases of the starburst,
and drop-out from the UV-selected samples.

%===========================================================================
\section{Conclusions}

We have studied the dependence of gas-phase metallicity 12+log(O/H) 
on stellar mass \mstar\ and SFR
on a few samples of galaxies from z=0 to z=3.3.
In the local universe, we find that 
metallicity is tightly related to both \mstar\ and SFR
(fig.~\ref{fig:massmet}), and this
defines the Fundamental Metallicity Relation (fig.~\ref{fig:cfr1}). 
The residual metallicity dispersion of local SDSS galaxies around 
this FMR is about 0.05~dex (fig.~\ref{fig:plotdif}), 
i.e, about 12\%. The well-known mass-metallicity relation, together with the 
luminosity-metallicity and velocity-metallicity relations, is one particular
projection of this relation into one plane, and neglecting the 
dependence of metallicity on SFR results in doubling the observed dispersion.

When high redshift galaxies are compared to the FMR defined locally, we find 
no evolution up to z=2.5, i.e., high-redshift galaxies follow
the same FMR defined by SDSS galaxies even if they have higher SFRs
(fig~\ref{fig:plotevol}).  
This means that the same physical processes are in place in the local universe
and at high redshifts. The observed evolution of the mass-metallicity relation 
is due to the increase of the average SFR with redshift, 
which results in sampling different parts
of the same FMR at different redshifts.

At even higher redshift, z$\sim$3.3,
evolution of $\sim$0.6~dex with respect to the FMR is found, although 
several observational effects and selection biases may affect the size of this 
evolution. This is an 
indication that different mechanisms start to dominate.

Even if the nature of the FMR is 3-dimensional in \mstar, SFR and 12+log(O/H),
metallicity is found to be tightly correlated with
$\mu_{0.32}=\rm{log}(M_\star)-0.32\,\rm{log}(\rm{SFR})$. 
Galaxies at any redshifts up to z=2.5
follow the same $\mu_{0.32}$-metallicity relation and have
the same range of values of $\mu_{0.32}$ (fig.\ref{fig:bestfit}).

Metallicity in galaxies of any mass is found to have the same dependence of SSFR,
with galaxies above the threshold of SSFR=$10^{-10}$yr$^{-1}$  showing a 
rapidly decreasing metallicities with increasing SSFR (fig.~\ref{fig:ssfrmet}).

The interpretation of the existence of the FMR, its dependence of SSFR, and the role of 
$\mu_{0.32}$ depend on the relevant timescales. If dynamical times are shorter than
timescales for chemical enrichment, the dependence of metallicity on SFR can be easily 
explained by
dilution by infall. In this case this effect dominates the metallicity evolution of 
galaxies at high redshift, when galaxies grow because of massive accretions of metal poor gas
and produce large SFR. Also galaxies at z=3.3 can fit into this scheme, with large 
masses of infalling gas.
This is in agreement with other recent independent 
results \citep{Mannucci09b,Cresci10}.
In the local universe, galaxies with large SFR are 
rare and often associated to merging events, and other effects
becomes dominant which relate metallicity mainly to \mstar. Outflow is an
possibly, although downsizing could also work. 
If, in contrast, infall and SFR evolve on timescales much longer than the 
chemical enrichment timescale,
a sort of steady-state situation is created: 
continuos infall of metal-poor gas, which both sustains SFR and dilutes metallicity,
and outflow of metal-rich gas in galactic winds. In this case the outflows must depend
on both  mass and SFR. 

The small residual scatter around the FMR in the local universe supports the 
smooth accretion 
scenario, where galaxy grow is dominated by continuos accretion of cold gas.
Interacting and merging galaxies are expected to show larger spread around the FMR,
in agreement to what is actually observed.
Galaxies at intermediate and high redshifts show 
larger metallicity dispersions, which could be due either to uncertainties 
in the measurements, or to intrinsic dispersion, or both. 
This effect prevents us to study the evolution of residual scatter
with redshift. Nevertheless, the absence of significant biasses in metallicity 
or in SFR up to z=2.5 points toward the existence of the same physical 
effects and the dominance of smooth accretion
even at intermediate redshift.\\

\noindent
{\bf Acknowledgements}

We are grateful to the MPA/JHU teams which made their measured quantities on SDSS galaxies
publicly available, and to the members of the italian Virtual Observatory for support 
on data visualization. We also acknowledge stimulating discussions with T. Nagao,
R. Dav\'e, S. Ellison, and the Arcetri MEGA group.

%===========================================================================
\bibliographystyle{/Users/filippo/arcetri/Papers/aa-package/bibtex/aa}
\bibliography{/Users/filippo/arcetri/bibdesk/Bibliography}

\end{document}